\documentclass[final,superscriptaddress,english,twocolumn,amssymb,aps,prx,
longbibliography]{revtex4-1}
\usepackage{graphicx}
\usepackage{natbib}
\usepackage{amsmath}
\usepackage{amssymb}
\usepackage{appendix}
\usepackage[dvipsnames]{xcolor}{\huge }
\usepackage[section]{placeins}
\definecolor{darkblue}{rgb}{0,0,.65}
\definecolor{darkgreen}{rgb}{0.28,0.41,0.19}

\usepackage{wasysym} 

\usepackage{nicefrac}
\usepackage[%
    pdfauthor={David J. Luitz},%
  pdfstartview=FitH,%
  breaklinks=true,%
  bookmarks=true,%
  colorlinks=true,%
  anchorcolor=black,%
  citecolor=blue,
  filecolor=black,%
  menucolor=black,%
  urlcolor=darkblue,%
  linkcolor=blue,%
 ]{hyperref}
\usepackage[all]{hypcap} %
\newcommand{\bra}[1]{\langle\,#1\,|}
\newcommand{\ket}[1]{|#1\rangle}

\graphicspath{{images/}}
\begin{document}

\title{Magnetization process and ordering of the $S=1/2$ pyrochlore Heisenberg antiferromagnet in a magnetic field}

\author{Imre Hagym\'asi}\email{hagymasi@pks.mpg.de}
\affiliation{Max Planck Institute for the Physics of Complex Systems, 
Noethnitzer Str. 38, 01187 Dresden, Germany}
\affiliation{Strongly Correlated Systems "Lend\"ulet" Research Group, Institute 
for Solid State
Physics and Optics, Wigner Research Centre for Physics, Budapest H-1525 P.O. 
Box 49, Hungary
}
\author{Robin Sch\"afer}\email{schaefer@pks.mpg.de}
\affiliation{Max Planck Institute for the Physics of Complex Systems, 
Noethnitzer Str. 38, 01187 Dresden, Germany}
\author{Roderich Moessner}\email{moessner@pks.mpg.de}
\author{David J. Luitz}
\email{dluitz@pks.mpg.de}
\affiliation{Max Planck Institute for the Physics of Complex Systems, 
Noethnitzer Str. 38, 01187 Dresden, Germany}

\date{\today}

\begin{abstract}
We study the $S=1/2$ pyrochlore Heisenberg antiferromagnet in a magnetic field. 
Using large scale density-matrix renormalization group (DMRG) calculations for clusters with up to $128$ spins, we find indications for a finite triplet gap, causing a threshold field to nonzero magnetization in the magnetization curve.
We obtain a robust saturation field consistent with a magnon crystal, although the corresponding $5/6$
magnetization plateau is very slim and possibly unstable. 
Most remarkably, there is a pronounced and apparently robust $1/ 2$
magnetization plateau where the ground state breaks the rotational symmetry of the lattice, exhibiting {\it oppositely polarized} spins on alternating kagom\'{e} and triangular planes. Reminiscent of the kagom\'{e} ice plateau of the pyrochlore Ising antiferromagnet known as spin ice, it arises via a much more subtle `quantum order-by-disorder' mechanism.
\end{abstract}
\maketitle

\emph{Introduction.---}
The Heisenberg antiferromagnet on the pyrochlore lattice, is one of the most frustrated three dimensional  magnets,  and as such a prime candidate for exotic, specifically quantum spin liquid, behavior. Indeed, its classical variants are topological magnets (classical spin liquids), exhibiting a Coulomb phase \cite{Isakov_dipolar_prl,Henley_dipolar_prb,Henley_ARCMP} in the limit of low temperature both for the Heisenberg \cite{moecha_pyro_prl} and the Ising (spin ice) variants, with the latter also hosting deconfined magnetic monopoles as quasiparticles
\cite{harris_geometrical_1997,bramwell_spin_2001,castelnovo_magnetic_2008,castelnovo_spin_2012}. 

Due to the complexity of the quantum $S=1/2$ problem and concomitant lack of unbiased methods, the nature of the ground state of the pyrochlore Heisenberg antiferromagnet is still being discussed \cite{harris_ordering_1991,tsunetsugu_prb_2001,Tsunetsugu_pyro_2001,isoda_valence_bond_1998,CanalsLacroix_prl,Berg_subcontractor_2003,kim_prb_2008,burnell_monopole_2009,iqbal_quantum_2019,smith_CeZrO_2021}, with recent progress indicating that the ground state breaks inversion symmetry \cite{hagymasi_prl_2021,schafer_pyrochlore_2020,astrakhantsev_broken-symmetry_2021} rather than being a quantum spin liquid. 

In a related strand of work, the study of frustrated magnets in an applied field \cite{Moessner_field} has turned up many interesting phenomena, such as string excitations and Kasteleyn transitions, dimensional reduction and much more. Particularly prominent has been the study of magnon crystals \cite{schollwoeck_qm_2004, honecker_magnon_2004, 
schnack_eigenstates_2006, derzhko_localized_magnons_2007, 
nishimoto_numerics_kagome_plateaus_2013,schulenburg_theory_kagome_magnon_groundstate_2002, honecker_kagome_2005} and magnetization plateau \cite{honecker_magnetization_2004,sakai_kagome_2011,capponi_numerics_kagome_plateaus_2013,capponi_numerical_2013,nakano_kagome_2014,schnack_kagome42_2018,plat_kagome_2018,pal_prb_2019,penc_prl_2004}, `incompressible' magnetic states, which may be stabilized when gaps between the ground states of different magnetization sectors remain finite in the thermodynamic limit.

\begin{figure}[t]
    \centering
    \includegraphics[width=0.49\columnwidth]{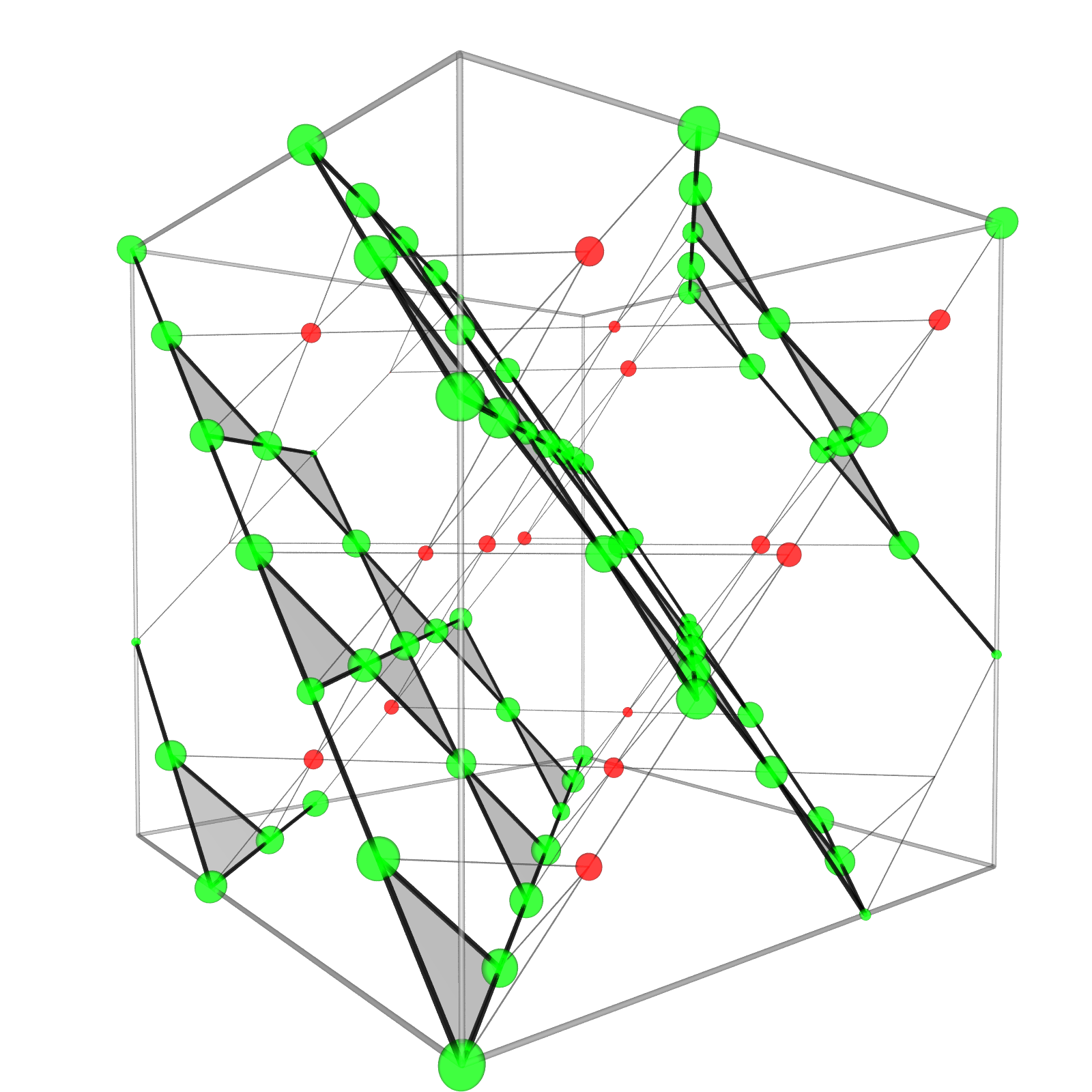}
     \includegraphics[width=0.49\columnwidth]{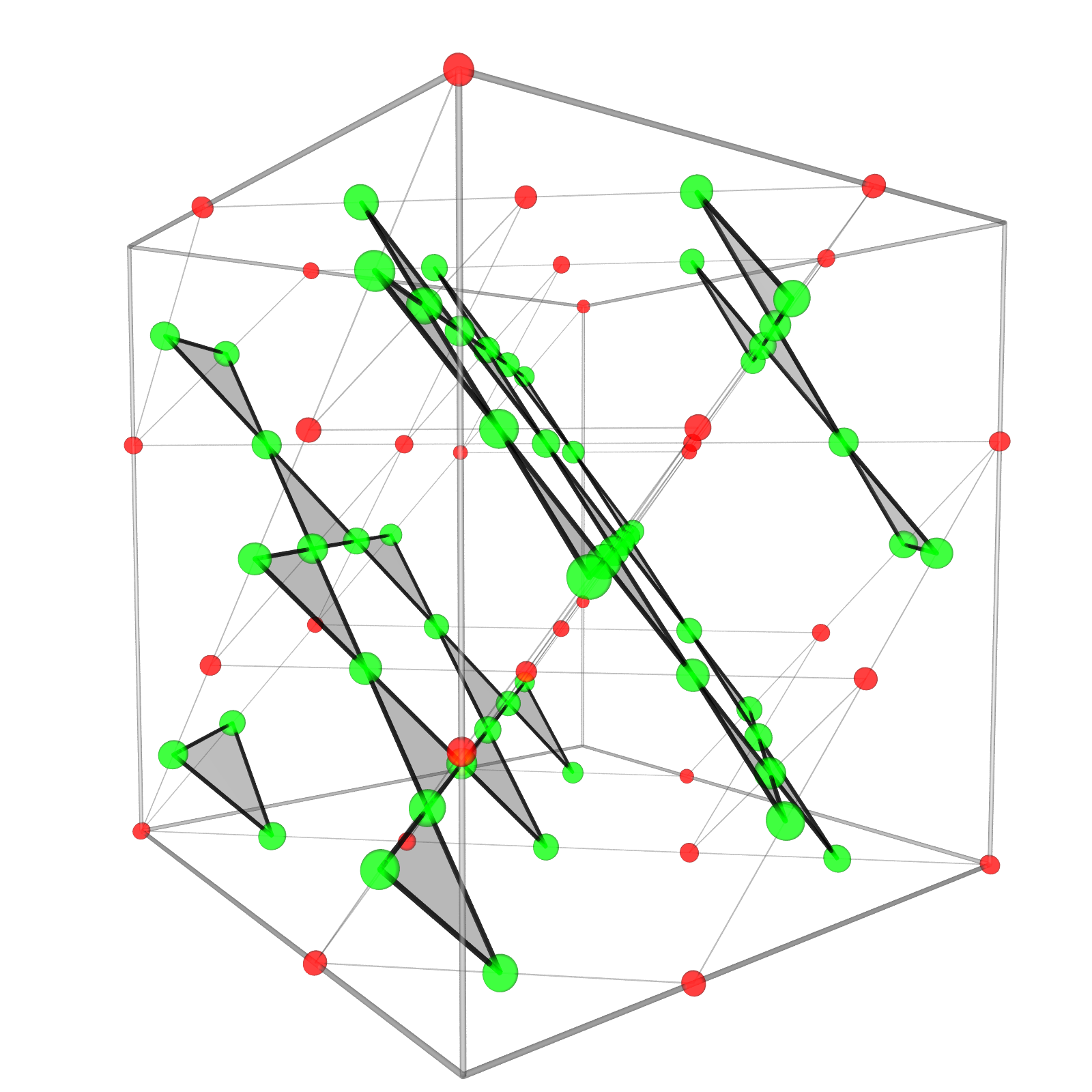}\\
    \caption{Onsite magnetization for the $m/m_s=1/2$ plateau of the 108 site 
and 128 site clusters. We show only the cubic unit cell cut of the clusters for better comparison. Green balls correspond to polarized spins in one direction, red balls represent a polarization in the opposite direction. The size of the balls is proportional to the onsite magnetization. Gray shaded triangles are guiding the eye to indicate the polarized kagom\'{e} planes.
}
        \label{fig:magnetization}
\end{figure}

The classical variants exhibit either no magnetization plateau  for the Heisenberg model (at least in the absence of magnetoelastic coupling \cite{penc_prl_2004,ueda_experiment_2005,kojima_experiment_2010}), or, in the Ising case, the very rich physics of kagom\'{e} ice \cite{matsuhira_new_2002,Udagawa_kagice,Moessner_kagice}. The latter arises for a field applied in a [111] direction, which has a large projection onto the local Ising axes on one quarter of the spins, which therefore quickly get polarised. The remaining three quarters of the spins reside on kagom\'{e} layers (green dots in Fig.~\ref{fig:magnetization}) which enter a stable partially polarised plateau at intermediate field strengths. 

In the quantum limit, this magnetization process remains largely unexplored \cite{pal_prb_2019,balents_pyrochlore_2006,zhitomirsky_pyrochlore_2007,zhitomirsky_pyrochlore_2000,coletta_frustrated_2013}, precisely due to the lack of reliable methods alluded to above, while the two-dimensional $S=1/2$ kagom\'{e} Heisenberg antiferromagnet has been thoroughly studied \cite{schnack_kagome42_2018,chen_kagome_2018,nakano_kagome_2018,schulenburg_theory_kagome_magnon_groundstate_2002,honecker_magnetization_2004,honecker_kagome_2005,sakai_kagome_2011,capponi_numerics_kagome_plateaus_2013,nishimoto_controlling_2013,capponi_numerical_2013,nakano_kagome_2014,plat_kagome_2018}. Although a material which can be modeled by an SU(2) symmetric $S = 1/2$ Heisenberg model is still lacking, higher spin Heisenberg materials are known, such as the $S = 1$ NaCaNi$_2$F$_7$ \cite{plumb_continuum_2019} or the $S = 3/2$ compound CdCr$_2$O$_4$. In the latter compound a robust 1/2 magnetization plateau has been observed \cite{ueda_experiment_2005}.

\begin{figure}[t]
    \centering
\includegraphics[width=\columnwidth]{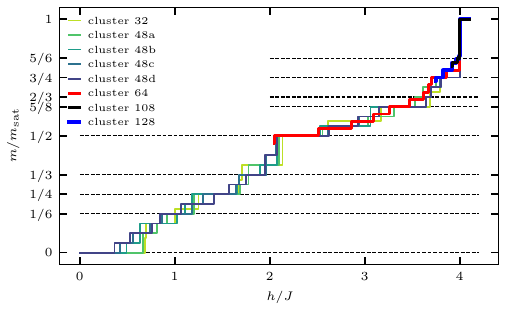}
\caption{Normalized ground state magnetization $m=\bra{\psi_0}S_\text{tot}^z\ket{\psi_0}$ of different pyrochlore clusters as a function of the external magnetic field $h$. The saturation magnetization is given by the fully polarized state of $N$ spins: $m_\text{sat}=N/2$.
For the 64 site cluster the upper half, while for the largest clusters 108 and 128, only the part of the curve at the strongest fields can be 
reliably calculated.}
        \label{fig:magnetization_curve_00}
\end{figure}

Here, we study the magnetization process of the $S=1/2$ Heisenberg antiferromagnet from zero field to saturation with DMRG, which has recently turned out to be very useful regarding the zero-field properties \cite{hagymasi_prl_2021}. Most saliently, we find an incompressible magnetic phase with a 3:1 spin polarization ratio, signaled by a robust plateau at half saturation.

The ground state corresponding to this plateau breaks the rotational symmetry of the lattice. It exhibits polarized kagom\'{e} planes along the field direction and {\it anti-}polarized interplane sites. Unlike in kagom\'{e} ice, this pattern of 3:1 disproportionation arises spontaneously, being selected from an exponentially large number of possible 3:1 disproportionations in what may be termed a quantum order by disorder mechanism. Also, the minority sublattice has {\it negative}, rather than enhanced, Zeeman energy -- a spontaneous instance of ferrimagnetism. 

Furthermore, we obtain a value of the saturation field which is consistent with what one obtains for the magnon crystal state, an exact eigenstate of a range of frustrated Heisenberg magnets \cite{schnack_eigenstates_2006}.

\emph{Methods.---} 
We determine the field-dependence of the ground state magnetization for periodic clusters ranging from $N=32$ to $N=128$ spins, using SU(2) and U(1) DMRG \cite{hubig:_syten_toolk,hubig17:_symmet_protec_tensor_networ,hubig_2015,McCulloch_2007}. Although DMRG is by nature a one dimensional method \cite{white_1992,white_1993,noack2005,schollwock_review_2011,hallberg_review}, it has been successfully used in two \cite{white_2d_dmrg} and recently three dimensions \cite{ummethum_numerics_2013,hagymasi_prl_2021}, by `linearizing' the system along a snake path at the price of non-local interactions within the snake. 
Conservation of the 
total spin, $S_{\text{tot}}$ and its $z$-component, $S^z_{\text{tot}}$ allows us to target and optimize the ground states within the concomitant symmetry sectors. 
We observe that despite the fact that the SU(2) representation is very efficient in higher spin sectors, the convergence while optimizing the energy is sometimes better when explicitly enforcing the U(1) spin symmetry only.
Since the wave function is represented as a matrix-product state with finite bond dimension,  
extrapolation to infinite bond dimension is necessary. This is usually done by either extrapolating as a function of the truncation error or the 
variance \cite{schollwock_review_2011}. 
We optimize the wave function for small bond dimensions ($\lesssim 2000$) using the two-site DMRG algorithm, but for larger bond dimensions we switch to the single-site variant with subspace expansion \cite{hubig_2015}. For the bond dimensions we used (up to $\sim 20000$ SU(2) or U(1) states) the calculation of the full variance is impractical, and we therefore extrapolate these energies to the error free limit as a function of the two-site variance. This quantity was shown to be equally good compared to the truncation error \cite{hubig_prb_2018}, see Appendix \ref{sec:numerical} for further details.

\begin{table}
	\centering
	\begin{tabular}{l@{\hspace{5mm}}c@{\hspace{5mm}}c@{\hspace{5mm}}c@{\hspace{5mm}}|c}
		\hline\hline
		cluster & $\vec{c}_1$ &  $\vec{c}_2$ & $\vec{c}_3$ & length  \\
		\hline
		32   &  $2 \vec{a}_1$ & $2 \vec{a}_2$ & $2\vec{a}_3$ & 4  \\
		48a   &  $(\frac{3}{2},\frac{1}{2},0)^T$ & $(0,1,1)^T$ & $(0,1,-1)^T$ & 4  \\
		48b   &  $(\frac{3}{2},\frac{1}{2},0)^T$ & $(0,\frac 1 2 ,\frac 3 2)^T$ & $(0,1,-1)^T$ & 4  \\
		48c   &  $(\frac{3}{2},1,\frac{1}{2})^T$ & $(0,1 ,-1)^T$ & $(1,-1,0)^T$ & 4  \\
		48d   &  $(1,1,1)^T$ & $(1 ,0,-1)^T$ & $(1,-1,0)^T$ & 4 \\
		64   &  $(1,1,1)^T$ & $(1 ,1,-1)^T$ & $(-1,1,1)^T$ & 6  \\
		108   &  $3 \vec{a}_1$ & $3 \vec{a}_2$ & $3\vec{a}_3$ & 6  \\
		128   &  $(2,0,0)^T$ & $(0 ,2,0)^T$ & $(0,0,2)^T$  & 8\\
		\hline
	\end{tabular}
	\caption{Cluster vectors $\vec{c}_1$,$\vec{c}_2$,$\vec{c}_3$ of the 8 clusters used in this work. The last column shows the length of the shortest loop winding across the periodic boundary and thus competing with the loops in the bulk (e.g. hexagons of length 6). The clusters of size $32$ and $108$ respect all lattice symmetries.}
	\label{tab:clusters}
\end{table}

\emph{Magnetization curve.--- }
We consider the ground state of the $S=1/2$ pyrochlore Heisenberg antiferromagnet
\begin{equation}
	H = J \sum_{\langle i, j\rangle} \vec{S}_i \cdot \vec{S}_j - h \sum_i S_i^z
\end{equation}
in a finite magnetic field $h>0$.
The spins are located on a pyrochlore lattice defined by the fcc lattice vectors $\vec{a}_1 = \frac{1}{2} (1,1,0)^T$, $\vec{a}_2 = \frac{1}{2} (1,0,1)^T$, $\vec{a}_3 = \frac{1}{2} (0,1,1)^T$, together with four tetrahedral basis vectors $\vec{b}_0=0$, $\vec{b_i} = \frac{1}{2} \vec{a}_i$.
This model conserves the total magnetization $S_\text{tot}^z = \sum_i S_i^z$, since $[H,S_\text{tot}^z]=0$, and hence the Hamiltonian decomposes into symmetry sectors with fixed total magnetization $m=-N/2,-N/2+1 \dots N/2$. This means that each eigenstate $\ket{n}_m$ of $H$ is also an eigenstate of $S_\text{tot}^z$: $S_\text{tot}^z \ket{n}_m = m \ket{n}_m$ and therefore the
eigenstates of the Hamiltonian are \emph{independent} of the field $h$, and their energy $E^n_{m}(h)$ is shifted with respect to the zero field value $E^n_{m}(0)$ by $E^n_{m}(h) = E^n_{m}(0) -h m$.

In the absence of a field, $h=0$, the ground state of the Hamiltonian is in the $m=0$ sector.
For finite values of $h>0$, the energies of states exhibiting a finite magnetization $m\neq 0$ will change by $-hm$ and potentially become the overall ground state of the system.
This leads to the characteristic jumps of the magnetization in Fig. \ref{fig:magnetization_curve_00}. The field strength at which a transition of the ground state magnetization from $m$ to $m+1$ occurs is entirely determined by the minimal energy of all states in the sectors $m$ and $m+1$ \emph{in the absence of the field}, $E_m^0(h=0) = \min_n E_m^n(h=0)$. The field where $E_{m+1}^0(h)$ becomes lower than $E_m^0(h)$ is determined by $E_{m+1}^0(h) = E_{m}^0(h)$, i.e.
\begin{equation}
    E_{m+1}^0(h=0) - h(m+1) = E_{m}^0(h=0) -h m,
\end{equation}
which is solved by $h= E_{m+1}^0(h=0) - E_{m}^0(h=0)$. Multiple transitions between $m$, $m+1$, $m+2\dots$ can coincide, leading to a larger change at a given field strength. 
To obtain the full magnetization curve for a given cluster, we therefore have to calculate the lowest energies in all magnetization sectors at zero field using DMRG, respecting the $U(1)$ symmetry associated with the conservation of $S_\text{tot}^z$. In fact, since the total spin $S^2_\text{tot}$ also commutes with both $H$ and $S_\text{tot}^z$, we can also use the full SU(2) symmetry as mentioned above.

If the ground states of adjacent $S_\text{tot}^z$ sectors 
are separated by a gap in the thermodynamic limit, the magnetization will remain locked into the lower magnetization sector for a range of fields proportional to the gap, leading to a plateau in the magnetization curve, and hence an incompressible magnetic state.

We display the resulting magnetization curves in Fig.~\ref{fig:magnetization_curve_00} for different clusters ranging from 32 to 128 sites, noting that for large clusters it is only possible to determine the large field part of the magnetization curve.

A prominent feature is the large jump of the magnetization from its maximum to $5/6\,m_{\rm sat}$ near the saturation field $h_{\text{sat}}=4J$.
Position and height of this jump can be understood using the same reasoning as in the case of kagom\'{e} 
\cite{schulenburg_theory_kagome_magnon_groundstate_2002,
capponi_numerics_kagome_plateaus_2013}. The exact ground state of symmetry 
sectors with very large $S_\text{tot}^z$ is a crystal of localized magnon modes, while the ground state of the sector with maximal $S_{\rm tot}^z$ is the trivial fully polarized state.
One possible densest packing of independent magnon modes on the pyrochlore lattice is 
given by densely arranging magnons localized on hexagonal motifs in the kagom\'{e} planes, leading to a 
magnetization plateau at $m/m_\text{sat}=5/6$ (see Appendix \ref{sec:magnon} for details). Each magnon mode reduces the energy by $4J$, so that $h_\text{sat}=4J$ (Eq. \eqref{eq:mag_02}). Each independent magnon excitation requires three unit cells (twelve sites), which fixes the densest packing. Due to the requirement of commensurability of the densest packing with the cluster geometry, we find the $5/6$ plateau only for the 108 site cluster in Fig. \ref{fig:magnetization_curve_00}. 
On other clusters we can obtain even larger jumps at $h=4J$, as modes localized on shorter loops winding across periodic boundary conditions can yield a denser magnon mode filling as a finite-size effect. This yields, for the 32 and 48 site clusters, a broad plateau at $m/m_\text{sat}=3/4$ which is not representative for the infinite lattice. The narrowness of the  $5/6$ plateau on the 108 site cluster in turn calls into question its stability in the thermodynamic limit.

\begin{figure}[!h]
    \centering
	\includegraphics[width=\columnwidth]{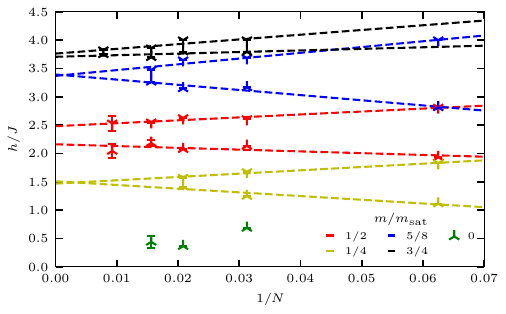}
	\caption{Extrapolation to the thermodynamic limit of magnetization jumps for the plateaus $m/m_\text{sat}=1/2$ (red), $5/8$ (blue) and $3/4$ (black) based on simple linear fits in $1/N$. The lower/higher ($h_-$/$h_+$) points of each plateau define the jump towards a lower/higher magnetization plateaus for different cluster sizes. We obtained the following values in the thermodynamic limit: $h_+^{1/2} = 2.48(1)J$, $h_-^{1/2} = 2.16(5)J$. To gain further data points for the finite-size scaling we included a cluster given by 4 unit cells (16 sites) with periodic boundary conditions.} 
        \label{fig:magnetization_width_00}
\end{figure}

We attempt to extrapolate the widths of the  magnetization plateau observed in finite size clusters (Fig. \ref{fig:magnetization_curve_00}) to the thermodynamic limit (Fig. \ref{fig:magnetization_width_00}), using linear fits in $1/N$. There is little indication that the $3/4$ and $5/8$ plateau retain a finite width. 
In contrast, we have strong evidence for a finite width of the $1/2$ plateau in the thermodynamic limit (red in Fig. \ref{fig:magnetization_width_00}), which is located in the field range $h_-^{1/2} \leq h \leq h_+^{1/2}$, with jumps at $h_-^{1/2} = 2.16(5)J$ and $h_+^{1/2} = 2.48(1)J$. 
The size of the $0$ plateau (determined by the triplet gap) varies non-monotonically with cluster size and geometry and is inaccessible for our largest clusters, rendering a smooth extrapolation impossible.
Nevertheless, our result for the triplet gap in the 64 site cluster, 0.42(11) agrees with that of the recent variational Monte Carlo estimate, 0.40(4), in the thermodynamic limit \cite{astrakhantsev_broken-symmetry_2021} and the data shown in Fig. \ref{fig:magnetization_width_00} is consistent with a finite triplet gap.

{It is worth commenting on the actual values of the magnetic field to make the comparison with experiments easier. Since a material, which realizes the $S=1/2$ Heisenberg model is still lacking, the closest relative we can consider is the $S = 1$ NaCaNi$_2$F$_7$ \cite{plumb_continuum_2019}. Assuming the same value of $J\sim 3.2 \ \mathrm{meV}$ in a $S=1/2$ material (and $g$ factor $g\sim2$), the saturation field would correspond to $B_{\mathrm{sat}}\sim 110 \ \mathrm{T}$ and the ${1}/{2}$ plateau is expected to start at $\sim 68 \ \mathrm{T}$. Such high values of magnetic fields are accessible in pulsed field experiments.}

\emph{Properties of the ${1}/{2}$ plateau.--- } We turn to the correlations of the  ${1}/{2}$ plateau.
We focus on the largest cluster, where finite-size effects due to short loops winding across periodic boundaries are suppressed. Nonetheless, we include results for smaller clusters for finite-size extrapolations, as in the preceding analysis. 
Indeed, while the 32 site cluster develops
a uniform magnetization throughout the lattice with $\bra{\psi_0}S^z_i\ket{\psi_0} \equiv0.25$, larger clusters exhibit a more complex pattern, stabilizing inequivalent spins with differing polarization. 
{These are arranged with respect to one of the four rotational axes defined within each tetrahedron}.
The planes perpendicular to this axis are alternating triangular and kagom\'{e} planes, containing 1/4 and 3/4 of the spins, shown in red and green in Fig.~\ref{fig:magnetization}, respectively: each tetrahedron contributes a 'base triangle' to the kagom\'{e}, and its apex spin to the triangular, plane.   

The kagom\'{e} spins (A sites in Tab. \ref{tab:averaged_onsite}) acquire a polarization along the field, while the triangular spins (B sites) are polarised {\it in the opposite direction} in our clusters of size 64, 108 and 128.
Fig. \ref{fig:magnetization} shows the onsite magnetization pattern $\bra{\psi_0} S_i^z \ket{\psi_0}$ in a cubic unit cell for the clusters of size 108 and 128, along with a  list for different clusters in Tab. \ref{tab:averaged_onsite}.
While the largest cluster with $128$ sites develops this pattern perfectly, smaller clusters can exhibit defects in the form of lines with small onsite magnetization passing through kagom\'{e} planes, which we attribute to the existence of short resonant loops across the periodic boundaries in these clusters. The number of such defective sites is listed in the caption of Tab. \ref{tab:averaged_onsite} and vanishes for the 128 cluster.

\begin{table}[h]
    \centering
    \begin{tabular}{l@{\hspace{8mm}}c@{\hspace{8mm}}c@{\hspace{8mm}}c}
        \hline\hline
         $N$ & $A$ & $B$  \\
        \hline
        $64$ &  $0.379 \pm 0.077$ & $-0.161 \pm 0.000$ \\\hline
       $108$ &  $0.419 \pm 0.036$ & $-0.245 \pm 0.044$ \\\hline
       $128$ &  $0.426 \pm 0.015$ & $-0.278 \pm 0.007$  \\\hline
    \end{tabular}
    \caption{Averaged onsite magnetization $\langle S^z_i\rangle$ and standard deviation across sites of the two types of sites $A$ and $B$ observed in the finite cluster with size $N$. We excluded the defects (9 of $N=64$, 6 of $N=108$ and 0 of $N=128$) in form of the lines with small onsite magnetization to determine the averaged values.}
    \label{tab:averaged_onsite}
\end{table}

We emphasize that this symmetry breaking is distinct from the one we found for the zero-field problem, where the lattice inversion symmetry appears to be broken, as the two sublattices of tetrahedrons have different energy densities~\cite{hagymasi_prl_2021}. 
On the 1/2 magnetization plateau, the inversion symmetry is preserved, while the rotational symmetry of the lattice is broken by the emergence of a preferred $[111]$ axis. The rotational symmetry around this axis is not broken, but those within the three other kagom\'{e} planes are.

\begin{figure}[t]
    \centering
    \includegraphics[width=\columnwidth]{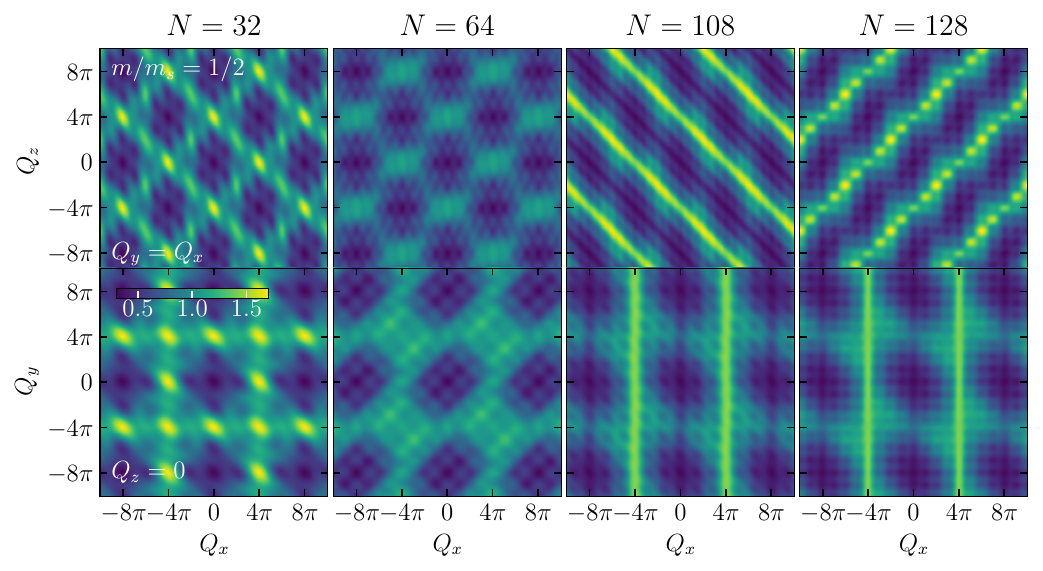}
    \caption{Spin structure factor $S(\vec{Q})$ for the $m/m_s=1/2$ plateau of various clusters. The top row shows the $Q_y=Q_x$ cut through the Brillouin zone, the bottom row corresponds to the $Q_z=0$ cut.}
        \label{fig:structure_factor_half_plateau}
\end{figure}

The symmetry breaking is naturally reflected in the spin structure factor 
\begin{equation}
    S(\vec{Q})= \frac{4}{3N} \sum_{i j} \langle \vec{S}_i\cdot 
\vec{S}_j\rangle_c \cos\left[\vec{Q}\cdot 
\left(\vec{R}_i-\vec{R}_j\right)\right],
\end{equation}
where $\vec{R}_i$ denote the real-space coordinates of sites and the index $c$ 
denotes the connected part of the correlation matrix (the factor $4/3$ comes from normalization 
$1/(S(S+1))$ for spin $S=1/2$). This is  plotted in 
Fig.~\ref{fig:structure_factor_half_plateau} for several  clusters. The larger clusters exhibit clearly discernible bright lines in the structure factor, discarding the rotational symmetry present for the 32-site cluster; in contrast to the $m/m_s=0$ state, the inversion symmetry remains intact. 

{Earlier work\cite{balents_pyrochlore_2006} predicts another possible pattern, the $R$-state, for the $S=3/2$ case. We discuss the competition of the $R$-state with the best variational wave function in Appendix \ref{sec:numerical} and conclude that the $R$-state has higher energy.}

\emph{Discussion ---}
The pyrochlore Heisenberg antiferromagnet in a field, like its zero-field cousin, illustrates the capacity of frustrated magnets to exhibit a plethora of instabilities, discarding the rotational symmetry at half-magnetization and forming an incompressible state. It is interesting to embed this finding in a broader context. 

Firstly, the idea that the half-magnetization plateau goes along with a 3:1 disproportionation of sites seems entirely natural: indeed, for collinear spins, such a 3:1 ratio is the only way to obtain half-magnetized tetrahedrons. Note, however, that such 3:1 tetrahedrons can be tiled in  exponentially numerous ways on the pyrochlore lattice, indeed mapping onto dimer coverings of the diamond lattice, which have a finite entropy of $S\approx0.13 k_B$ \cite{vanderstraeten_residual_2018,nagle_lattice_1966,nagle_lattice_1966_diamond_lattice}. 
The selection through fluctuations of a specific one (or subset) of such tiling with lowered symmetry is known as order-by-disorder. In this sense, our magnetization plateau exhibits a form of quantum order-by-disorder, although the assignment of what term of the Hamiltonian contributes the fluctuations is to some degree a matter of choice. At any rate, the emergence of ordered ferrimagnetism in the plateau is a striking instability of a highly  frustrated quantum magnet.  

We close by contrasting the $1/2$ plateau to the kagom\'{e} ice plateau mentioned in the introduction. In a real pyrochlore material with spin-orbit coupling, an Ising anisotropy needs to go along with non-collinear easy axes: the local easy axis is the $[111]$ direction joining a site with the centers of its tetrahedrons. This turns a uniform applied magnetic field into a staggered Zeeman field according to the sublattice \cite{Moessner_relief}; as mentioned above, when applied along a $[111]$ direction, it polarizes the triangular planes more strongly than the kagom\'{e} ones, as the easy axes projections differ by a factor of 3. We find it most intriguing that this general setting arises for the $S=1/2$ Heisenberg plateau by spontaneous rather than explicit symmetry breaking, and the question of interpolating between the two immediately poses itself. 
Note that the two cases differ considerably in (non-symmetry) `details': the triangular layers are strongly positively polarised in kagom\'{e} ice, in contrast to their weak {\it negative} polarization in the Heisenberg $S=1/2$ plateau. 

There clearly remains much further scope for studying the highly frustrated quantum magnets in $d=3$ with and without applied fields, and for the foreseeable future, recent technological progress promises to yield previously inaccessible interesting data such as those underpinning the present article.  

\begin{acknowledgments}
We acknowledge  support from the Deutsche 
Forschungsgemeinschaft through SFB 1143 (project-id 247310070) and cluster of excellence 
ct.qmat (EXC 2147, project-id 390858490).
I.H.~was supported in part by the Hungarian National Research,   
Development   and   Innovation Office (NKFIH) through Grants No.~K120569 and No.~K134983. Some of the data presented here was produced using the \textsc{SyTen} toolkit \cite{hubig:_syten_toolk,hubig17:_symmet_protec_tensor_networ}.
\end{acknowledgments}

\bibliography{pyrochlore,pyrochlore_field}

\begin{thebibliography}{69}%
\makeatletter
\providecommand \@ifxundefined [1]{%
 \@ifx{#1\undefined}
}%
\providecommand \@ifnum [1]{%
 \ifnum #1\expandafter \@firstoftwo
 \else \expandafter \@secondoftwo
 \fi
}%
\providecommand \@ifx [1]{%
 \ifx #1\expandafter \@firstoftwo
 \else \expandafter \@secondoftwo
 \fi
}%
\providecommand \natexlab [1]{#1}%
\providecommand \enquote  [1]{``#1''}%
\providecommand \bibnamefont  [1]{#1}%
\providecommand \bibfnamefont [1]{#1}%
\providecommand \citenamefont [1]{#1}%
\providecommand \href@noop [0]{\@secondoftwo}%
\providecommand \href [0]{\begingroup \@sanitize@url \@href}%
\providecommand \@href[1]{\@@startlink{#1}\@@href}%
\providecommand \@@href[1]{\endgroup#1\@@endlink}%
\providecommand \@sanitize@url [0]{\catcode `\\12\catcode `\$12\catcode
  `\&12\catcode `\#12\catcode `\^12\catcode `\_12\catcode `\%12\relax}%
\providecommand \@@startlink[1]{}%
\providecommand \@@endlink[0]{}%
\providecommand \url  [0]{\begingroup\@sanitize@url \@url }%
\providecommand \@url [1]{\endgroup\@href {#1}{\urlprefix }}%
\providecommand \urlprefix  [0]{URL }%
\providecommand \Eprint [0]{\href }%
\providecommand \doibase [0]{http://dx.doi.org/}%
\providecommand \selectlanguage [0]{\@gobble}%
\providecommand \bibinfo  [0]{\@secondoftwo}%
\providecommand \bibfield  [0]{\@secondoftwo}%
\providecommand \translation [1]{[#1]}%
\providecommand \BibitemOpen [0]{}%
\providecommand \bibitemStop [0]{}%
\providecommand \bibitemNoStop [0]{.\EOS\space}%
\providecommand \EOS [0]{\spacefactor3000\relax}%
\providecommand \BibitemShut  [1]{\csname bibitem#1\endcsname}%
\let\auto@bib@innerbib\@empty
\bibitem [{\citenamefont {{Isakov}}\ \emph {et~al.}(2004)\citenamefont
  {{Isakov}}, \citenamefont {{Gregor}}, \citenamefont {{Moessner}},\ and\
  \citenamefont {{Sondhi}}}]{Isakov_dipolar_prl}%
  \BibitemOpen
  \bibfield  {author} {\bibinfo {author} {\bibfnamefont {S.~V.}\ \bibnamefont
  {{Isakov}}}, \bibinfo {author} {\bibfnamefont {K.}~\bibnamefont {{Gregor}}},
  \bibinfo {author} {\bibfnamefont {R.}~\bibnamefont {{Moessner}}}, \ and\
  \bibinfo {author} {\bibfnamefont {S.~L.}\ \bibnamefont {{Sondhi}}},\
  }\bibfield  {title} {\enquote {\bibinfo {title} {{Dipolar Spin Correlations
  in Classical Pyrochlore Magnets}},}\ }\href {\doibase
  10.1103/PhysRevLett.93.167204} {\bibfield  {journal} {\bibinfo  {journal}
  {\prl}\ }\textbf {\bibinfo {volume} {93}},\ \bibinfo {eid} {167204} (\bibinfo
  {year} {2004})}\BibitemShut {NoStop}%
\bibitem [{\citenamefont {Henley}(2005)}]{Henley_dipolar_prb}%
  \BibitemOpen
  \bibfield  {author} {\bibinfo {author} {\bibfnamefont {C.~L.}\ \bibnamefont
  {Henley}},\ }\bibfield  {title} {\enquote {\bibinfo {title} {Power-law spin
  correlations in pyrochlore antiferromagnets},}\ }\href {\doibase
  10.1103/PhysRevB.71.014424} {\bibfield  {journal} {\bibinfo  {journal} {Phys.
  Rev. B}\ }\textbf {\bibinfo {volume} {71}},\ \bibinfo {pages} {014424}
  (\bibinfo {year} {2005})}\BibitemShut {NoStop}%
\bibitem [{\citenamefont {{Henley}}(2010)}]{Henley_ARCMP}%
  \BibitemOpen
  \bibfield  {author} {\bibinfo {author} {\bibfnamefont {Christopher~L.}\
  \bibnamefont {{Henley}}},\ }\bibfield  {title} {\enquote {\bibinfo {title}
  {{The ``Coulomb Phase'' in Frustrated Systems}},}\ }\href {\doibase
  10.1146/annurev-conmatphys-070909-104138} {\bibfield  {journal} {\bibinfo
  {journal} {Annual Review of Condensed Matter Physics}\ }\textbf {\bibinfo
  {volume} {1}},\ \bibinfo {pages} {179--210} (\bibinfo {year}
  {2010})}\BibitemShut {NoStop}%
\bibitem [{\citenamefont {Moessner}\ and\ \citenamefont
  {Chalker}(1998)}]{moecha_pyro_prl}%
  \BibitemOpen
  \bibfield  {author} {\bibinfo {author} {\bibfnamefont {R.}~\bibnamefont
  {Moessner}}\ and\ \bibinfo {author} {\bibfnamefont {J.~T.}\ \bibnamefont
  {Chalker}},\ }\bibfield  {title} {\enquote {\bibinfo {title} {Properties of a
  classical spin liquid: The heisenberg pyrochlore antiferromagnet},}\ }\href
  {\doibase 10.1103/PhysRevLett.80.2929} {\bibfield  {journal} {\bibinfo
  {journal} {Phys. Rev. Lett.}\ }\textbf {\bibinfo {volume} {80}},\ \bibinfo
  {pages} {2929--2932} (\bibinfo {year} {1998})}\BibitemShut {NoStop}%
\bibitem [{\citenamefont {Harris}\ \emph {et~al.}(1997)\citenamefont {Harris},
  \citenamefont {Bramwell}, \citenamefont {McMorrow}, \citenamefont {Zeiske},\
  and\ \citenamefont {Godfrey}}]{harris_geometrical_1997}%
  \BibitemOpen
  \bibfield  {author} {\bibinfo {author} {\bibfnamefont {M.~J.}\ \bibnamefont
  {Harris}}, \bibinfo {author} {\bibfnamefont {S.~T.}\ \bibnamefont
  {Bramwell}}, \bibinfo {author} {\bibfnamefont {D.~F.}\ \bibnamefont
  {McMorrow}}, \bibinfo {author} {\bibfnamefont {T.}~\bibnamefont {Zeiske}}, \
  and\ \bibinfo {author} {\bibfnamefont {K.~W.}\ \bibnamefont {Godfrey}},\
  }\bibfield  {title} {\enquote {\bibinfo {title} {Geometrical {Frustration} in
  the {Ferromagnetic} {Pyrochlore} {Ho$_2$Ti$_2$O$_7$}},}\ }\href {\doibase
  10.1103/PhysRevLett.79.2554} {\bibfield  {journal} {\bibinfo  {journal}
  {Physical Review Letters}\ }\textbf {\bibinfo {volume} {79}},\ \bibinfo
  {pages} {2554--2557} (\bibinfo {year} {1997})}\BibitemShut {NoStop}%
\bibitem [{\citenamefont {Bramwell}\ and\ \citenamefont
  {Gingras}(2001)}]{bramwell_spin_2001}%
  \BibitemOpen
  \bibfield  {author} {\bibinfo {author} {\bibfnamefont {Steven~T.}\
  \bibnamefont {Bramwell}}\ and\ \bibinfo {author} {\bibfnamefont {Michel
  J.~P.}\ \bibnamefont {Gingras}},\ }\bibfield  {title} {\enquote {\bibinfo
  {title} {Spin {Ice} {State} in {Frustrated} {Magnetic} {Pyrochlore}
  {Materials}},}\ }\href {\doibase 10.1126/science.1064761} {\bibfield
  {journal} {\bibinfo  {journal} {Science}\ }\textbf {\bibinfo {volume}
  {294}},\ \bibinfo {pages} {1495--1501} (\bibinfo {year} {2001})}\BibitemShut
  {NoStop}%
\bibitem [{\citenamefont {Castelnovo}\ \emph {et~al.}(2008)\citenamefont
  {Castelnovo}, \citenamefont {Moessner},\ and\ \citenamefont
  {Sondhi}}]{castelnovo_magnetic_2008}%
  \BibitemOpen
  \bibfield  {author} {\bibinfo {author} {\bibfnamefont {C.}~\bibnamefont
  {Castelnovo}}, \bibinfo {author} {\bibfnamefont {R.}~\bibnamefont
  {Moessner}}, \ and\ \bibinfo {author} {\bibfnamefont {S.~L.}\ \bibnamefont
  {Sondhi}},\ }\bibfield  {title} {\enquote {\bibinfo {title} {Magnetic
  monopoles in spin ice},}\ }\href {\doibase 10.1038/nature06433} {\bibfield
  {journal} {\bibinfo  {journal} {Nature}\ }\textbf {\bibinfo {volume} {451}},\
  \bibinfo {pages} {42--45} (\bibinfo {year} {2008})}\BibitemShut {NoStop}%
\bibitem [{\citenamefont {Castelnovo}\ \emph {et~al.}(2012)\citenamefont
  {Castelnovo}, \citenamefont {Moessner},\ and\ \citenamefont
  {Sondhi}}]{castelnovo_spin_2012}%
  \BibitemOpen
  \bibfield  {author} {\bibinfo {author} {\bibfnamefont {C.}~\bibnamefont
  {Castelnovo}}, \bibinfo {author} {\bibfnamefont {R.}~\bibnamefont
  {Moessner}}, \ and\ \bibinfo {author} {\bibfnamefont {S.L.}\ \bibnamefont
  {Sondhi}},\ }\bibfield  {title} {\enquote {\bibinfo {title} {Spin {Ice},
  {Fractionalization}, and {Topological} {Order}},}\ }\href {\doibase
  10.1146/annurev-conmatphys-020911-125058} {\bibfield  {journal} {\bibinfo
  {journal} {Annual Review of Condensed Matter Physics}\ }\textbf {\bibinfo
  {volume} {3}},\ \bibinfo {pages} {35--55} (\bibinfo {year}
  {2012})}\BibitemShut {NoStop}%
\bibitem [{\citenamefont {Harris}\ \emph {et~al.}(1991)\citenamefont {Harris},
  \citenamefont {Berlinsky},\ and\ \citenamefont
  {Bruder}}]{harris_ordering_1991}%
  \BibitemOpen
  \bibfield  {author} {\bibinfo {author} {\bibfnamefont {A.~B.}\ \bibnamefont
  {Harris}}, \bibinfo {author} {\bibfnamefont {A.~J.}\ \bibnamefont
  {Berlinsky}}, \ and\ \bibinfo {author} {\bibfnamefont {C.}~\bibnamefont
  {Bruder}},\ }\bibfield  {title} {\enquote {\bibinfo {title} {Ordering by
  quantum fluctuations in a strongly frustrated {Heisenberg}
  antiferromagnet},}\ }\href {\doibase 10.1063/1.348098} {\bibfield  {journal}
  {\bibinfo  {journal} {Journal of Applied Physics}\ }\textbf {\bibinfo
  {volume} {69}},\ \bibinfo {pages} {5200} (\bibinfo {year}
  {1991})}\BibitemShut {NoStop}%
\bibitem [{\citenamefont
  {Tsunetsugu}(2001{\natexlab{a}})}]{tsunetsugu_prb_2001}%
  \BibitemOpen
  \bibfield  {author} {\bibinfo {author} {\bibfnamefont {Hirokazu}\
  \bibnamefont {Tsunetsugu}},\ }\bibfield  {title} {\enquote {\bibinfo {title}
  {Spin-singlet order in a pyrochlore antiferromagnet},}\ }\href {\doibase
  10.1103/PhysRevB.65.024415} {\bibfield  {journal} {\bibinfo  {journal} {Phys.
  Rev. B}\ }\textbf {\bibinfo {volume} {65}},\ \bibinfo {pages} {024415}
  (\bibinfo {year} {2001}{\natexlab{a}})}\BibitemShut {NoStop}%
\bibitem [{\citenamefont
  {Tsunetsugu}(2001{\natexlab{b}})}]{Tsunetsugu_pyro_2001}%
  \BibitemOpen
  \bibfield  {author} {\bibinfo {author} {\bibfnamefont {Hirokazu}\
  \bibnamefont {Tsunetsugu}},\ }\bibfield  {title} {\enquote {\bibinfo {title}
  {Antiferromagnetic quantum spins on the pyrochlore lattice},}\ }\href
  {\doibase 10.1143/jpsj.70.640} {\bibfield  {journal} {\bibinfo  {journal}
  {Journal of the Physical Society of Japan}\ }\textbf {\bibinfo {volume}
  {70}},\ \bibinfo {pages} {640–643} (\bibinfo {year}
  {2001}{\natexlab{b}})}\BibitemShut {NoStop}%
\bibitem [{\citenamefont {Isoda}\ and\ \citenamefont
  {Mori}(1998)}]{isoda_valence_bond_1998}%
  \BibitemOpen
  \bibfield  {author} {\bibinfo {author} {\bibfnamefont {Makoto}\ \bibnamefont
  {Isoda}}\ and\ \bibinfo {author} {\bibfnamefont {Shigeyoshi}\ \bibnamefont
  {Mori}},\ }\bibfield  {title} {\enquote {\bibinfo {title} {Valence-{Bond}
  {Crystal} and {Anisotropic} {Excitation} {Spectrum} on 3-{Dimensionally}
  {Frustrated} {Pyrochlore}},}\ }\href {\doibase 10.1143/JPSJ.67.4022}
  {\bibfield  {journal} {\bibinfo  {journal} {Journal of the Physical Society
  of Japan}\ }\textbf {\bibinfo {volume} {67}},\ \bibinfo {pages} {4022--4025}
  (\bibinfo {year} {1998})}\BibitemShut {NoStop}%
\bibitem [{\citenamefont {Canals}\ and\ \citenamefont
  {Lacroix}(1998)}]{CanalsLacroix_prl}%
  \BibitemOpen
  \bibfield  {author} {\bibinfo {author} {\bibfnamefont {B.}~\bibnamefont
  {Canals}}\ and\ \bibinfo {author} {\bibfnamefont {C.}~\bibnamefont
  {Lacroix}},\ }\bibfield  {title} {\enquote {\bibinfo {title} {Pyrochlore
  antiferromagnet: A three-dimensional quantum spin liquid},}\ }\href {\doibase
  10.1103/PhysRevLett.80.2933} {\bibfield  {journal} {\bibinfo  {journal}
  {Phys. Rev. Lett.}\ }\textbf {\bibinfo {volume} {80}},\ \bibinfo {pages}
  {2933--2936} (\bibinfo {year} {1998})}\BibitemShut {NoStop}%
\bibitem [{\citenamefont {Berg}\ \emph {et~al.}(2003)\citenamefont {Berg},
  \citenamefont {Altman},\ and\ \citenamefont
  {Auerbach}}]{Berg_subcontractor_2003}%
  \BibitemOpen
  \bibfield  {author} {\bibinfo {author} {\bibfnamefont {Erez}\ \bibnamefont
  {Berg}}, \bibinfo {author} {\bibfnamefont {Ehud}\ \bibnamefont {Altman}}, \
  and\ \bibinfo {author} {\bibfnamefont {Assa}\ \bibnamefont {Auerbach}},\
  }\bibfield  {title} {\enquote {\bibinfo {title} {Singlet excitations in
  pyrochlore: A study of quantum frustration},}\ }\href
  {http://dx.doi.org/10.1103/PhysRevLett.90.147204} {\bibfield  {journal}
  {\bibinfo  {journal} {Physical Review Letters}\ }\textbf {\bibinfo {volume}
  {90}} (\bibinfo {year} {2003})}\BibitemShut {NoStop}%
\bibitem [{\citenamefont {Kim}\ and\ \citenamefont {Han}(2008)}]{kim_prb_2008}%
  \BibitemOpen
  \bibfield  {author} {\bibinfo {author} {\bibfnamefont {Jung~Hoon}\
  \bibnamefont {Kim}}\ and\ \bibinfo {author} {\bibfnamefont {Jung~Hoon}\
  \bibnamefont {Han}},\ }\bibfield  {title} {\enquote {\bibinfo {title} {Chiral
  spin states in the pyrochlore heisenberg magnet: Fermionic mean-field theory
  and variational monte carlo calculations},}\ }\href {\doibase
  10.1103/PhysRevB.78.180410} {\bibfield  {journal} {\bibinfo  {journal} {Phys.
  Rev. B}\ }\textbf {\bibinfo {volume} {78}},\ \bibinfo {pages} {180410}
  (\bibinfo {year} {2008})}\BibitemShut {NoStop}%
\bibitem [{\citenamefont {Burnell}\ \emph {et~al.}(2009)\citenamefont
  {Burnell}, \citenamefont {Chakravarty},\ and\ \citenamefont
  {Sondhi}}]{burnell_monopole_2009}%
  \BibitemOpen
  \bibfield  {author} {\bibinfo {author} {\bibfnamefont {F.~J.}\ \bibnamefont
  {Burnell}}, \bibinfo {author} {\bibfnamefont {Shoibal}\ \bibnamefont
  {Chakravarty}}, \ and\ \bibinfo {author} {\bibfnamefont {S.~L.}\ \bibnamefont
  {Sondhi}},\ }\bibfield  {title} {\enquote {\bibinfo {title} {Monopole flux
  state on the pyrochlore lattice},}\ }\href {\doibase
  10.1103/PhysRevB.79.144432} {\bibfield  {journal} {\bibinfo  {journal}
  {Physical Review B}\ }\textbf {\bibinfo {volume} {79}},\ \bibinfo {pages}
  {144432} (\bibinfo {year} {2009})}\BibitemShut {NoStop}%
\bibitem [{\citenamefont {Iqbal}\ \emph {et~al.}(2019)\citenamefont {Iqbal},
  \citenamefont {M{\"u}ller}, \citenamefont {Ghosh}, \citenamefont {Gingras},
  \citenamefont {Jeschke}, \citenamefont {Rachel}, \citenamefont {Reuther},\
  and\ \citenamefont {Thomale}}]{iqbal_quantum_2019}%
  \BibitemOpen
  \bibfield  {author} {\bibinfo {author} {\bibfnamefont {Yasir}\ \bibnamefont
  {Iqbal}}, \bibinfo {author} {\bibfnamefont {Tobias}\ \bibnamefont
  {M{\"u}ller}}, \bibinfo {author} {\bibfnamefont {Pratyay}\ \bibnamefont
  {Ghosh}}, \bibinfo {author} {\bibfnamefont {Michel J.~P.}\ \bibnamefont
  {Gingras}}, \bibinfo {author} {\bibfnamefont {Harald~O.}\ \bibnamefont
  {Jeschke}}, \bibinfo {author} {\bibfnamefont {Stephan}\ \bibnamefont
  {Rachel}}, \bibinfo {author} {\bibfnamefont {Johannes}\ \bibnamefont
  {Reuther}}, \ and\ \bibinfo {author} {\bibfnamefont {Ronny}\ \bibnamefont
  {Thomale}},\ }\bibfield  {title} {\enquote {\bibinfo {title} {Quantum and
  {Classical} {Phases} of the {Pyrochlore} {Heisenberg} {Model} with
  {Competing} {Interactions}},}\ }\href {\doibase 10.1103/PhysRevX.9.011005}
  {\bibfield  {journal} {\bibinfo  {journal} {Phys. Rev. X}\ }\textbf {\bibinfo
  {volume} {9}},\ \bibinfo {pages} {011005} (\bibinfo {year}
  {2019})}\BibitemShut {NoStop}%
\bibitem [{\citenamefont {Smith}\ \emph {et~al.}(2022)\citenamefont {Smith},
  \citenamefont {Benton}, \citenamefont {Yahne}, \citenamefont {Placke},
  \citenamefont {Sch\"afer}, \citenamefont {Gaudet}, \citenamefont {Dudemaine},
  \citenamefont {Fitterman}, \citenamefont {Beare}, \citenamefont {Wildes},
  \citenamefont {Bhattacharya}, \citenamefont {DeLazzer}, \citenamefont
  {Buhariwalla}, \citenamefont {Butch}, \citenamefont {Movshovich},
  \citenamefont {Garrett}, \citenamefont {Marjerrison}, \citenamefont {Clancy},
  \citenamefont {Kermarrec}, \citenamefont {Luke}, \citenamefont {Bianchi},
  \citenamefont {Ross},\ and\ \citenamefont {Gaulin}}]{smith_CeZrO_2021}%
  \BibitemOpen
  \bibfield  {author} {\bibinfo {author} {\bibfnamefont {E.~M.}\ \bibnamefont
  {Smith}}, \bibinfo {author} {\bibfnamefont {O.}~\bibnamefont {Benton}},
  \bibinfo {author} {\bibfnamefont {D.~R.}\ \bibnamefont {Yahne}}, \bibinfo
  {author} {\bibfnamefont {B.}~\bibnamefont {Placke}}, \bibinfo {author}
  {\bibfnamefont {R.}~\bibnamefont {Sch\"afer}}, \bibinfo {author}
  {\bibfnamefont {J.}~\bibnamefont {Gaudet}}, \bibinfo {author} {\bibfnamefont
  {J.}~\bibnamefont {Dudemaine}}, \bibinfo {author} {\bibfnamefont
  {A.}~\bibnamefont {Fitterman}}, \bibinfo {author} {\bibfnamefont
  {J.}~\bibnamefont {Beare}}, \bibinfo {author} {\bibfnamefont {A.~R.}\
  \bibnamefont {Wildes}}, \bibinfo {author} {\bibfnamefont {S.}~\bibnamefont
  {Bhattacharya}}, \bibinfo {author} {\bibfnamefont {T.}~\bibnamefont
  {DeLazzer}}, \bibinfo {author} {\bibfnamefont {C.~R.~C.}\ \bibnamefont
  {Buhariwalla}}, \bibinfo {author} {\bibfnamefont {N.~P.}\ \bibnamefont
  {Butch}}, \bibinfo {author} {\bibfnamefont {R.}~\bibnamefont {Movshovich}},
  \bibinfo {author} {\bibfnamefont {J.~D.}\ \bibnamefont {Garrett}}, \bibinfo
  {author} {\bibfnamefont {C.~A.}\ \bibnamefont {Marjerrison}}, \bibinfo
  {author} {\bibfnamefont {J.~P.}\ \bibnamefont {Clancy}}, \bibinfo {author}
  {\bibfnamefont {E.}~\bibnamefont {Kermarrec}}, \bibinfo {author}
  {\bibfnamefont {G.~M.}\ \bibnamefont {Luke}}, \bibinfo {author}
  {\bibfnamefont {A.~D.}\ \bibnamefont {Bianchi}}, \bibinfo {author}
  {\bibfnamefont {K.~A.}\ \bibnamefont {Ross}}, \ and\ \bibinfo {author}
  {\bibfnamefont {B.~D.}\ \bibnamefont {Gaulin}},\ }\bibfield  {title}
  {\enquote {\bibinfo {title} {Case for a $\text{U}(1)_\pi$ {Quantum} {Spin}
  {Liquid} {Ground} {State} in the {Dipole}-{Octupole} {Pyrochlore}
  $\text{Ce}_{2}\text{Zr}_{2}\text{O}_{7}$},}\ }\href {\doibase
  10.1103/PhysRevX.12.021015} {\bibfield  {journal} {\bibinfo  {journal}
  {Physical Review X}\ }\textbf {\bibinfo {volume} {12}},\ \bibinfo {pages}
  {021015} (\bibinfo {year} {2022})}\BibitemShut {NoStop}%
\bibitem [{\citenamefont {Hagym\'asi}\ \emph {et~al.}(2021)\citenamefont
  {Hagym\'asi}, \citenamefont {Sch\"afer}, \citenamefont {Moessner},\ and\
  \citenamefont {Luitz}}]{hagymasi_prl_2021}%
  \BibitemOpen
  \bibfield  {author} {\bibinfo {author} {\bibfnamefont {Imre}\ \bibnamefont
  {Hagym\'asi}}, \bibinfo {author} {\bibfnamefont {Robin}\ \bibnamefont
  {Sch\"afer}}, \bibinfo {author} {\bibfnamefont {Roderich}\ \bibnamefont
  {Moessner}}, \ and\ \bibinfo {author} {\bibfnamefont {David~J.}\ \bibnamefont
  {Luitz}},\ }\bibfield  {title} {\enquote {\bibinfo {title} {Possible
  inversion symmetry breaking in the s=1/2 pyrochlore heisenberg magnet},}\
  }\href {\doibase 10.1103/PhysRevLett.126.117204} {\bibfield  {journal}
  {\bibinfo  {journal} {Phys. Rev. Lett.}\ }\textbf {\bibinfo {volume} {126}},\
  \bibinfo {pages} {117204} (\bibinfo {year} {2021})}\BibitemShut {NoStop}%
\bibitem [{\citenamefont {Sch\"afer}\ \emph {et~al.}(2020)\citenamefont
  {Sch\"afer}, \citenamefont {Hagymási}, \citenamefont {Moessner},\ and\
  \citenamefont {Luitz}}]{schafer_pyrochlore_2020}%
  \BibitemOpen
  \bibfield  {author} {\bibinfo {author} {\bibfnamefont {Robin}\ \bibnamefont
  {Sch\"afer}}, \bibinfo {author} {\bibfnamefont {Imre}\ \bibnamefont
  {Hagymási}}, \bibinfo {author} {\bibfnamefont {Roderich}\ \bibnamefont
  {Moessner}}, \ and\ \bibinfo {author} {\bibfnamefont {David~J.}\ \bibnamefont
  {Luitz}},\ }\bibfield  {title} {\enquote {\bibinfo {title} {Pyrochlore
  $\text{S}=\frac{1}{2}$ {Heisenberg} antiferromagnet at finite temperature},}\
  }\href {\doibase 10.1103/PhysRevB.102.054408} {\bibfield  {journal} {\bibinfo
   {journal} {Physical Review B}\ }\textbf {\bibinfo {volume} {102}},\ \bibinfo
  {pages} {054408} (\bibinfo {year} {2020})}\BibitemShut {NoStop}%
\bibitem [{\citenamefont {Astrakhantsev}\ \emph {et~al.}(2021)\citenamefont
  {Astrakhantsev}, \citenamefont {Westerhout}, \citenamefont {Tiwari},
  \citenamefont {Choo}, \citenamefont {Chen}, \citenamefont {Fischer},
  \citenamefont {Carleo},\ and\ \citenamefont
  {Neupert}}]{astrakhantsev_broken-symmetry_2021}%
  \BibitemOpen
  \bibfield  {author} {\bibinfo {author} {\bibfnamefont {Nikita}\ \bibnamefont
  {Astrakhantsev}}, \bibinfo {author} {\bibfnamefont {Tom}\ \bibnamefont
  {Westerhout}}, \bibinfo {author} {\bibfnamefont {Apoorv}\ \bibnamefont
  {Tiwari}}, \bibinfo {author} {\bibfnamefont {Kenny}\ \bibnamefont {Choo}},
  \bibinfo {author} {\bibfnamefont {Ao}~\bibnamefont {Chen}}, \bibinfo {author}
  {\bibfnamefont {Mark~H.}\ \bibnamefont {Fischer}}, \bibinfo {author}
  {\bibfnamefont {Giuseppe}\ \bibnamefont {Carleo}}, \ and\ \bibinfo {author}
  {\bibfnamefont {Titus}\ \bibnamefont {Neupert}},\ }\bibfield  {title}
  {\enquote {\bibinfo {title} {Broken-{Symmetry} {Ground} {States} of the
  {Heisenberg} {Model} on the {Pyrochlore} {Lattice}},}\ }\href {\doibase
  10.1103/PhysRevX.11.041021} {\bibfield  {journal} {\bibinfo  {journal}
  {Physical Review X}\ }\textbf {\bibinfo {volume} {11}},\ \bibinfo {pages}
  {041021} (\bibinfo {year} {2021})}\BibitemShut {NoStop}%
\bibitem [{\citenamefont {{Moessner}}(2009)}]{Moessner_field}%
  \BibitemOpen
  \bibfield  {author} {\bibinfo {author} {\bibfnamefont {Roderich}\
  \bibnamefont {{Moessner}}},\ }\bibfield  {title} {\enquote {\bibinfo {title}
  {{Unconventional magnets in external magnetic fields}},}\ }in\ \href
  {\doibase 10.1088/1742-6596/145/1/012001} {\emph {\bibinfo {booktitle}
  {Journal of Physics Conference Series}}},\ \bibinfo {series} {Journal of
  Physics Conference Series}, Vol.\ \bibinfo {volume} {145}\ (\bibinfo {year}
  {2009})\ p.\ \bibinfo {pages} {012001}\BibitemShut {NoStop}%
\bibitem [{\citenamefont {Schollw{\"o}ck}\ \emph {et~al.}(2004)\citenamefont
  {Schollw{\"o}ck}, \citenamefont {Richter}, \citenamefont {Farnell},\ and\
  \citenamefont {Bishop}}]{schollwoeck_qm_2004}%
  \BibitemOpen
  \bibinfo {editor} {\bibfnamefont {Ulrich}\ \bibnamefont {Schollw{\"o}ck}},
  \bibinfo {editor} {\bibfnamefont {Johannes}\ \bibnamefont {Richter}},
  \bibinfo {editor} {\bibfnamefont {Damian J.~J.}\ \bibnamefont {Farnell}}, \
  and\ \bibinfo {editor} {\bibfnamefont {Raymod~F.}\ \bibnamefont {Bishop}},\
  eds.,\ \href {\doibase 10.1007/b96825} {\emph {\bibinfo {title} {Quantum
  {Magnetism}}}},\ Vol.\ \bibinfo {volume} {645}\ (\bibinfo  {publisher}
  {Springer},\ \bibinfo {address} {Berlin, Heidelberg},\ \bibinfo {year}
  {2004})\BibitemShut {NoStop}%
\bibitem [{\citenamefont {Honecker}\ \emph
  {et~al.}(2004{\natexlab{a}})\citenamefont {Honecker}, \citenamefont
  {Schulenburg},\ and\ \citenamefont {Richter}}]{honecker_magnon_2004}%
  \BibitemOpen
  \bibfield  {author} {\bibinfo {author} {\bibfnamefont {A}~\bibnamefont
  {Honecker}}, \bibinfo {author} {\bibfnamefont {J}~\bibnamefont
  {Schulenburg}}, \ and\ \bibinfo {author} {\bibfnamefont {J}~\bibnamefont
  {Richter}},\ }\bibfield  {title} {\enquote {\bibinfo {title} {Magnetization
  plateaus in frustrated antiferromagnetic quantum spin models},}\ }\href
  {\doibase 10.1088/0953-8984/16/11/025} {\bibfield  {journal} {\bibinfo
  {journal} {Journal of Physics: Condensed Matter}\ }\textbf {\bibinfo {volume}
  {16}},\ \bibinfo {pages} {S749--S758} (\bibinfo {year}
  {2004}{\natexlab{a}})}\BibitemShut {NoStop}%
\bibitem [{\citenamefont {Schnack}\ \emph {et~al.}(2006)\citenamefont
  {Schnack}, \citenamefont {Schmidt}, \citenamefont {Honecker}, \citenamefont
  {Schulenburg},\ and\ \citenamefont {Richter}}]{schnack_eigenstates_2006}%
  \BibitemOpen
  \bibfield  {author} {\bibinfo {author} {\bibfnamefont {J}~\bibnamefont
  {Schnack}}, \bibinfo {author} {\bibfnamefont {H-J}\ \bibnamefont {Schmidt}},
  \bibinfo {author} {\bibfnamefont {A}~\bibnamefont {Honecker}}, \bibinfo
  {author} {\bibfnamefont {J}~\bibnamefont {Schulenburg}}, \ and\ \bibinfo
  {author} {\bibfnamefont {J}~\bibnamefont {Richter}},\ }\bibfield  {title}
  {\enquote {\bibinfo {title} {Exact eigenstates of highly frustrated spin
  lattices probed in high fields},}\ }\href {\doibase
  10.1088/1742-6596/51/1/007} {\bibfield  {journal} {\bibinfo  {journal}
  {Journal of Physics: Conference Series}\ }\textbf {\bibinfo {volume} {51}},\
  \bibinfo {pages} {43--46} (\bibinfo {year} {2006})}\BibitemShut {NoStop}%
\bibitem [{\citenamefont {Derzhko}\ \emph {et~al.}(2007)\citenamefont
  {Derzhko}, \citenamefont {Richter}, \citenamefont {Honecker},\ and\
  \citenamefont {Schmidt}}]{derzhko_localized_magnons_2007}%
  \BibitemOpen
  \bibfield  {author} {\bibinfo {author} {\bibfnamefont {O.}~\bibnamefont
  {Derzhko}}, \bibinfo {author} {\bibfnamefont {J.}~\bibnamefont {Richter}},
  \bibinfo {author} {\bibfnamefont {A.}~\bibnamefont {Honecker}}, \ and\
  \bibinfo {author} {\bibfnamefont {H.-J.}\ \bibnamefont {Schmidt}},\
  }\bibfield  {title} {\enquote {\bibinfo {title} {Universal properties of
  highly frustrated quantum magnets in strong magnetic fields},}\ }\href
  {\doibase 10.1063/1.2780166} {\bibfield  {journal} {\bibinfo  {journal} {Low
  Temperature Physics}\ }\textbf {\bibinfo {volume} {33}},\ \bibinfo {pages}
  {745--756} (\bibinfo {year} {2007})}\BibitemShut {NoStop}%
\bibitem [{\citenamefont {Nishimoto}\ \emph
  {et~al.}(2013{\natexlab{a}})\citenamefont {Nishimoto}, \citenamefont
  {Shibata},\ and\ \citenamefont
  {Hotta}}]{nishimoto_numerics_kagome_plateaus_2013}%
  \BibitemOpen
  \bibfield  {author} {\bibinfo {author} {\bibfnamefont {Satoshi}\ \bibnamefont
  {Nishimoto}}, \bibinfo {author} {\bibfnamefont {Naokazu}\ \bibnamefont
  {Shibata}}, \ and\ \bibinfo {author} {\bibfnamefont {Chisa}\ \bibnamefont
  {Hotta}},\ }\bibfield  {title} {\enquote {\bibinfo {title} {Controlling
  frustrated liquids and solids with an applied field in a kagome heisenberg
  antiferromagnet},}\ }\href {\doibase 10.1038/ncomms3287} {\bibfield
  {journal} {\bibinfo  {journal} {Nature Communications}\ }\textbf {\bibinfo
  {volume} {4}},\ \bibinfo {pages} {2287} (\bibinfo {year}
  {2013}{\natexlab{a}})}\BibitemShut {NoStop}%
\bibitem [{\citenamefont {Schulenburg}\ \emph {et~al.}(2002)\citenamefont
  {Schulenburg}, \citenamefont {Honecker}, \citenamefont {Schnack},
  \citenamefont {Richter},\ and\ \citenamefont
  {Schmidt}}]{schulenburg_theory_kagome_magnon_groundstate_2002}%
  \BibitemOpen
  \bibfield  {author} {\bibinfo {author} {\bibfnamefont {J.}~\bibnamefont
  {Schulenburg}}, \bibinfo {author} {\bibfnamefont {A.}~\bibnamefont
  {Honecker}}, \bibinfo {author} {\bibfnamefont {J.}~\bibnamefont {Schnack}},
  \bibinfo {author} {\bibfnamefont {J.}~\bibnamefont {Richter}}, \ and\
  \bibinfo {author} {\bibfnamefont {H.-J.}\ \bibnamefont {Schmidt}},\
  }\bibfield  {title} {\enquote {\bibinfo {title} {Macroscopic magnetization
  jumps due to independent magnons in frustrated quantum spin lattices},}\
  }\href {\doibase 10.1103/PhysRevLett.88.167207} {\bibfield  {journal}
  {\bibinfo  {journal} {Phys. Rev. Lett.}\ }\textbf {\bibinfo {volume} {88}},\
  \bibinfo {pages} {167207} (\bibinfo {year} {2002})}\BibitemShut {NoStop}%
\bibitem [{\citenamefont {Honecker}\ \emph {et~al.}(2005)\citenamefont
  {Honecker}, \citenamefont {Cabra}, \citenamefont {Grynberg}, \citenamefont
  {Holdsworth}, \citenamefont {Pujol}, \citenamefont {Richter}, \citenamefont
  {Schmalfu{\ss}},\ and\ \citenamefont {Schulenburg}}]{honecker_kagome_2005}%
  \BibitemOpen
  \bibfield  {author} {\bibinfo {author} {\bibfnamefont {A.}~\bibnamefont
  {Honecker}}, \bibinfo {author} {\bibfnamefont {D.C.}\ \bibnamefont {Cabra}},
  \bibinfo {author} {\bibfnamefont {M.D.}\ \bibnamefont {Grynberg}}, \bibinfo
  {author} {\bibfnamefont {P.C.W.}\ \bibnamefont {Holdsworth}}, \bibinfo
  {author} {\bibfnamefont {P.}~\bibnamefont {Pujol}}, \bibinfo {author}
  {\bibfnamefont {J.}~\bibnamefont {Richter}}, \bibinfo {author} {\bibfnamefont
  {D.}~\bibnamefont {Schmalfu{\ss}}}, \ and\ \bibinfo {author} {\bibfnamefont
  {J.}~\bibnamefont {Schulenburg}},\ }\bibfield  {title} {\enquote {\bibinfo
  {title} {Ground state and low-lying excitations of the spin-$\frac{1}{2}$
  $\text{XXZ}$ model on the kagom{\'e} lattice at magnetization
  $\frac{1}{3}$},}\ }\href {\doibase
  https://doi.org/10.1016/j.physb.2005.01.430} {\bibfield  {journal} {\bibinfo
  {journal} {Physica B: Condensed Matter}\ }\textbf {\bibinfo {volume}
  {359-361}},\ \bibinfo {pages} {1391--1393} (\bibinfo {year}
  {2005})}\BibitemShut {NoStop}%
\bibitem [{\citenamefont {Honecker}\ \emph
  {et~al.}(2004{\natexlab{b}})\citenamefont {Honecker}, \citenamefont
  {Schulenburg},\ and\ \citenamefont {Richter}}]{honecker_magnetization_2004}%
  \BibitemOpen
  \bibfield  {author} {\bibinfo {author} {\bibfnamefont {A.}~\bibnamefont
  {Honecker}}, \bibinfo {author} {\bibfnamefont {J.}~\bibnamefont
  {Schulenburg}}, \ and\ \bibinfo {author} {\bibfnamefont {J.}~\bibnamefont
  {Richter}},\ }\bibfield  {title} {\enquote {\bibinfo {title} {Magnetization
  plateaus in frustrated antiferromagnetic quantum spin models},}\ }\href
  {\doibase 10.1088/0953-8984/16/11/025} {\bibfield  {journal} {\bibinfo
  {journal} {J. Phys.: Condens. Matter}\ }\textbf {\bibinfo {volume} {16}},\
  \bibinfo {pages} {S749--S758} (\bibinfo {year}
  {2004}{\natexlab{b}})}\BibitemShut {NoStop}%
\bibitem [{\citenamefont {Sakai}\ and\ \citenamefont
  {Nakano}(2011)}]{sakai_kagome_2011}%
  \BibitemOpen
  \bibfield  {author} {\bibinfo {author} {\bibfnamefont {T\^oru}\ \bibnamefont
  {Sakai}}\ and\ \bibinfo {author} {\bibfnamefont {Hiroki}\ \bibnamefont
  {Nakano}},\ }\bibfield  {title} {\enquote {\bibinfo {title} {Critical
  magnetization behavior of the triangular- and kagome-lattice quantum
  antiferromagnets},}\ }\href {\doibase 10.1103/PhysRevB.83.100405} {\bibfield
  {journal} {\bibinfo  {journal} {Phys. Rev. B}\ }\textbf {\bibinfo {volume}
  {83}},\ \bibinfo {pages} {100405} (\bibinfo {year} {2011})}\BibitemShut
  {NoStop}%
\bibitem [{\citenamefont {Capponi}\ \emph {et~al.}(2013)\citenamefont
  {Capponi}, \citenamefont {Derzhko}, \citenamefont {Honecker}, \citenamefont
  {L\"auchli},\ and\ \citenamefont
  {Richter}}]{capponi_numerics_kagome_plateaus_2013}%
  \BibitemOpen
  \bibfield  {author} {\bibinfo {author} {\bibfnamefont {Sylvain}\ \bibnamefont
  {Capponi}}, \bibinfo {author} {\bibfnamefont {Oleg}\ \bibnamefont {Derzhko}},
  \bibinfo {author} {\bibfnamefont {Andreas}\ \bibnamefont {Honecker}},
  \bibinfo {author} {\bibfnamefont {Andreas~M.}\ \bibnamefont {L\"auchli}}, \
  and\ \bibinfo {author} {\bibfnamefont {Johannes}\ \bibnamefont {Richter}},\
  }\bibfield  {title} {\enquote {\bibinfo {title} {Numerical study of
  magnetization plateaus in the spin-$\frac{1}{2}$ kagome heisenberg
  antiferromagnet},}\ }\href {\doibase 10.1103/PhysRevB.88.144416} {\bibfield
  {journal} {\bibinfo  {journal} {Phys. Rev. B}\ }\textbf {\bibinfo {volume}
  {88}},\ \bibinfo {pages} {144416} (\bibinfo {year} {2013})}\BibitemShut
  {NoStop}%
\bibitem [{\citenamefont {Capponi}(2017)}]{capponi_numerical_2013}%
  \BibitemOpen
  \bibfield  {author} {\bibinfo {author} {\bibfnamefont {Sylvain}\ \bibnamefont
  {Capponi}},\ }\bibfield  {title} {\enquote {\bibinfo {title} {Numerical study
  of magnetization plateaus in the spin-$\frac{1}{2}$ heisenberg
  antiferromagnet on the checkerboard lattice},}\ }\href {\doibase
  10.1103/PhysRevB.95.014420} {\bibfield  {journal} {\bibinfo  {journal} {Phys.
  Rev. B}\ }\textbf {\bibinfo {volume} {95}},\ \bibinfo {pages} {014420}
  (\bibinfo {year} {2017})}\BibitemShut {NoStop}%
\bibitem [{\citenamefont {Nakano}\ and\ \citenamefont
  {Sakai}(2014)}]{nakano_kagome_2014}%
  \BibitemOpen
  \bibfield  {author} {\bibinfo {author} {\bibfnamefont {Hiroki}\ \bibnamefont
  {Nakano}}\ and\ \bibinfo {author} {\bibfnamefont {Tôru}\ \bibnamefont
  {Sakai}},\ }\bibfield  {title} {\enquote {\bibinfo {title} {Anomalous
  behavior of the magnetization process of the s = 1/2 kagome-lattice
  heisenberg antiferromagnet at one-third height of the saturation},}\ }\href
  {\doibase 10.7566/JPSJ.83.104710} {\bibfield  {journal} {\bibinfo  {journal}
  {Journal of the Physical Society of Japan}\ }\textbf {\bibinfo {volume}
  {83}},\ \bibinfo {pages} {104710} (\bibinfo {year} {2014})}\BibitemShut
  {NoStop}%
\bibitem [{\citenamefont {Schnack}\ \emph {et~al.}(2018)\citenamefont
  {Schnack}, \citenamefont {Schulenburg},\ and\ \citenamefont
  {Richter}}]{schnack_kagome42_2018}%
  \BibitemOpen
  \bibfield  {author} {\bibinfo {author} {\bibfnamefont {J\"urgen}\
  \bibnamefont {Schnack}}, \bibinfo {author} {\bibfnamefont {J\"org}\
  \bibnamefont {Schulenburg}}, \ and\ \bibinfo {author} {\bibfnamefont
  {Johannes}\ \bibnamefont {Richter}},\ }\bibfield  {title} {\enquote {\bibinfo
  {title} {Magnetism of the $n=42$ kagome lattice antiferromagnet},}\ }\href
  {\doibase 10.1103/PhysRevB.98.094423} {\bibfield  {journal} {\bibinfo
  {journal} {Phys. Rev. B}\ }\textbf {\bibinfo {volume} {98}},\ \bibinfo
  {pages} {094423} (\bibinfo {year} {2018})}\BibitemShut {NoStop}%
\bibitem [{\citenamefont {Plat}\ \emph {et~al.}(2018)\citenamefont {Plat},
  \citenamefont {Momoi},\ and\ \citenamefont {Hotta}}]{plat_kagome_2018}%
  \BibitemOpen
  \bibfield  {author} {\bibinfo {author} {\bibfnamefont {Xavier}\ \bibnamefont
  {Plat}}, \bibinfo {author} {\bibfnamefont {Tsutomu}\ \bibnamefont {Momoi}}, \
  and\ \bibinfo {author} {\bibfnamefont {Chisa}\ \bibnamefont {Hotta}},\
  }\bibfield  {title} {\enquote {\bibinfo {title} {Kinetic frustration induced
  supersolid in the $s=\frac{1}{2}$ kagome lattice antiferromagnet in a
  magnetic field},}\ }\href {\doibase 10.1103/PhysRevB.98.014415} {\bibfield
  {journal} {\bibinfo  {journal} {Phys. Rev. B}\ }\textbf {\bibinfo {volume}
  {98}},\ \bibinfo {pages} {014415} (\bibinfo {year} {2018})}\BibitemShut
  {NoStop}%
\bibitem [{\citenamefont {Pal}\ and\ \citenamefont {Lal}(2019)}]{pal_prb_2019}%
  \BibitemOpen
  \bibfield  {author} {\bibinfo {author} {\bibfnamefont {Santanu}\ \bibnamefont
  {Pal}}\ and\ \bibinfo {author} {\bibfnamefont {Siddhartha}\ \bibnamefont
  {Lal}},\ }\bibfield  {title} {\enquote {\bibinfo {title} {Magnetization
  plateaus of the quantum pyrochlore heisenberg antiferromagnet},}\ }\href
  {\doibase 10.1103/PhysRevB.100.104421} {\bibfield  {journal} {\bibinfo
  {journal} {Phys. Rev. B}\ }\textbf {\bibinfo {volume} {100}},\ \bibinfo
  {pages} {104421} (\bibinfo {year} {2019})}\BibitemShut {NoStop}%
\bibitem [{\citenamefont {Penc}\ \emph {et~al.}(2004)\citenamefont {Penc},
  \citenamefont {Shannon},\ and\ \citenamefont {Shiba}}]{penc_prl_2004}%
  \BibitemOpen
  \bibfield  {author} {\bibinfo {author} {\bibfnamefont {Karlo}\ \bibnamefont
  {Penc}}, \bibinfo {author} {\bibfnamefont {Nic}\ \bibnamefont {Shannon}}, \
  and\ \bibinfo {author} {\bibfnamefont {Hiroyuki}\ \bibnamefont {Shiba}},\
  }\bibfield  {title} {\enquote {\bibinfo {title} {Half-magnetization plateau
  stabilized by structural distortion in the antiferromagnetic heisenberg model
  on a pyrochlore lattice},}\ }\href {\doibase 10.1103/PhysRevLett.93.197203}
  {\bibfield  {journal} {\bibinfo  {journal} {Phys. Rev. Lett.}\ }\textbf
  {\bibinfo {volume} {93}},\ \bibinfo {pages} {197203} (\bibinfo {year}
  {2004})}\BibitemShut {NoStop}%
\bibitem [{\citenamefont {Ueda}\ \emph {et~al.}(2005)\citenamefont {Ueda},
  \citenamefont {Katori}, \citenamefont {Mitamura}, \citenamefont {Goto},\ and\
  \citenamefont {Takagi}}]{ueda_experiment_2005}%
  \BibitemOpen
  \bibfield  {author} {\bibinfo {author} {\bibfnamefont {Hiroaki}\ \bibnamefont
  {Ueda}}, \bibinfo {author} {\bibfnamefont {Hiroko~Aruga}\ \bibnamefont
  {Katori}}, \bibinfo {author} {\bibfnamefont {Hiroyuki}\ \bibnamefont
  {Mitamura}}, \bibinfo {author} {\bibfnamefont {Tsuneaki}\ \bibnamefont
  {Goto}}, \ and\ \bibinfo {author} {\bibfnamefont {Hidenori}\ \bibnamefont
  {Takagi}},\ }\bibfield  {title} {\enquote {\bibinfo {title} {Magnetic-field
  induced transition to the $1/2$ magnetization plateau state in the
  geometrically frustrated magnet $\text{CdCr}_2\text{O}_4$},}\ }\href
  {\doibase 10.1103/PhysRevLett.94.047202} {\bibfield  {journal} {\bibinfo
  {journal} {Phys. Rev. Lett.}\ }\textbf {\bibinfo {volume} {94}},\ \bibinfo
  {pages} {047202} (\bibinfo {year} {2005})}\BibitemShut {NoStop}%
\bibitem [{\citenamefont {Kojima}\ \emph {et~al.}(2010)\citenamefont {Kojima},
  \citenamefont {Miyata}, \citenamefont {Motome}, \citenamefont {Ueda},
  \citenamefont {Ueda},\ and\ \citenamefont
  {Takeyama}}]{kojima_experiment_2010}%
  \BibitemOpen
  \bibfield  {author} {\bibinfo {author} {\bibfnamefont {E.}~\bibnamefont
  {Kojima}}, \bibinfo {author} {\bibfnamefont {A.}~\bibnamefont {Miyata}},
  \bibinfo {author} {\bibfnamefont {Y.}~\bibnamefont {Motome}}, \bibinfo
  {author} {\bibfnamefont {H.}~\bibnamefont {Ueda}}, \bibinfo {author}
  {\bibfnamefont {Y.}~\bibnamefont {Ueda}}, \ and\ \bibinfo {author}
  {\bibfnamefont {S.}~\bibnamefont {Takeyama}},\ }\bibfield  {title} {\enquote
  {\bibinfo {title} {Magnetic orders of highly frustrated spinel,
  $\text{ZnCr}_2\text{O}_4$ in magnetic fields up to 400 $\text{T}$},}\ }\href
  {\doibase 10.1007/s10909-009-0069-7} {\bibfield  {journal} {\bibinfo
  {journal} {Journal of Low Temperature Physics}\ }\textbf {\bibinfo {volume}
  {159}},\ \bibinfo {pages} {3--6} (\bibinfo {year} {2010})}\BibitemShut
  {NoStop}%
\bibitem [{\citenamefont {Matsuhira}\ \emph {et~al.}(2002)\citenamefont
  {Matsuhira}, \citenamefont {Hiroi}, \citenamefont {Tayama}, \citenamefont
  {Takagi},\ and\ \citenamefont {Sakakibara}}]{matsuhira_new_2002}%
  \BibitemOpen
  \bibfield  {author} {\bibinfo {author} {\bibfnamefont {K.}~\bibnamefont
  {Matsuhira}}, \bibinfo {author} {\bibfnamefont {Z.}~\bibnamefont {Hiroi}},
  \bibinfo {author} {\bibfnamefont {T.}~\bibnamefont {Tayama}}, \bibinfo
  {author} {\bibfnamefont {S.}~\bibnamefont {Takagi}}, \ and\ \bibinfo {author}
  {\bibfnamefont {T.}~\bibnamefont {Sakakibara}},\ }\bibfield  {title}
  {\enquote {\bibinfo {title} {A new macroscopically degenerate ground state in
  the spin ice compound $\text{Dy}_2\text{Ti}_2\text{O}_7$ under a magnetic
  field},}\ }\href {\doibase 10.1088/0953-8984/14/29/101} {\bibfield  {journal}
  {\bibinfo  {journal} {Journal of Physics: Condensed Matter}\ }\textbf
  {\bibinfo {volume} {14}},\ \bibinfo {pages} {L559--L565} (\bibinfo {year}
  {2002})}\BibitemShut {NoStop}%
\bibitem [{\citenamefont {{Udagawa}}\ \emph {et~al.}(2002)\citenamefont
  {{Udagawa}}, \citenamefont {{Ogata}},\ and\ \citenamefont
  {{Hiroi}}}]{Udagawa_kagice}%
  \BibitemOpen
  \bibfield  {author} {\bibinfo {author} {\bibfnamefont {Masafumi}\
  \bibnamefont {{Udagawa}}}, \bibinfo {author} {\bibfnamefont {Masao}\
  \bibnamefont {{Ogata}}}, \ and\ \bibinfo {author} {\bibfnamefont {Zenji}\
  \bibnamefont {{Hiroi}}},\ }\bibfield  {title} {\enquote {\bibinfo {title}
  {{Exact Result of Ground-State Entropy for Ising Pyrochlore Magnets under a
  Magnetic Field along [111] Axis}},}\ }\href {\doibase 10.1143/JPSJ.71.2365}
  {\bibfield  {journal} {\bibinfo  {journal} {Journal of the Physical Society
  of Japan}\ }\textbf {\bibinfo {volume} {71}},\ \bibinfo {pages} {2365}
  (\bibinfo {year} {2002})}\BibitemShut {NoStop}%
\bibitem [{\citenamefont {{Moessner}}\ and\ \citenamefont
  {{Sondhi}}(2003)}]{Moessner_kagice}%
  \BibitemOpen
  \bibfield  {author} {\bibinfo {author} {\bibfnamefont {R.}~\bibnamefont
  {{Moessner}}}\ and\ \bibinfo {author} {\bibfnamefont {S.~L.}\ \bibnamefont
  {{Sondhi}}},\ }\bibfield  {title} {\enquote {\bibinfo {title} {{Theory of the
  [111] magnetization plateau in spin ice}},}\ }\href {\doibase
  10.1103/PhysRevB.68.064411} {\bibfield  {journal} {\bibinfo  {journal}
  {\prb}\ }\textbf {\bibinfo {volume} {68}},\ \bibinfo {eid} {064411} (\bibinfo
  {year} {2003})}\BibitemShut {NoStop}%
\bibitem [{\citenamefont {Bergman}\ \emph {et~al.}(2006)\citenamefont
  {Bergman}, \citenamefont {Shindou}, \citenamefont {Fiete},\ and\
  \citenamefont {Balents}}]{balents_pyrochlore_2006}%
  \BibitemOpen
  \bibfield  {author} {\bibinfo {author} {\bibfnamefont {Doron~L.}\
  \bibnamefont {Bergman}}, \bibinfo {author} {\bibfnamefont {Ryuichi}\
  \bibnamefont {Shindou}}, \bibinfo {author} {\bibfnamefont {Gregory~A.}\
  \bibnamefont {Fiete}}, \ and\ \bibinfo {author} {\bibfnamefont {Leon}\
  \bibnamefont {Balents}},\ }\bibfield  {title} {\enquote {\bibinfo {title}
  {Quantum effects in a half-polarized pyrochlore antiferromagnet},}\ }\href
  {\doibase 10.1103/PhysRevLett.96.097207} {\bibfield  {journal} {\bibinfo
  {journal} {Phys. Rev. Lett.}\ }\textbf {\bibinfo {volume} {96}},\ \bibinfo
  {pages} {097207} (\bibinfo {year} {2006})}\BibitemShut {NoStop}%
\bibitem [{\citenamefont {Zhitomirsky}\ and\ \citenamefont
  {Tsunetsugu}(2007)}]{zhitomirsky_pyrochlore_2007}%
  \BibitemOpen
  \bibfield  {author} {\bibinfo {author} {\bibfnamefont {M.~E.}\ \bibnamefont
  {Zhitomirsky}}\ and\ \bibinfo {author} {\bibfnamefont {Hirokazu}\
  \bibnamefont {Tsunetsugu}},\ }\bibfield  {title} {\enquote {\bibinfo {title}
  {Lattice gas description of pyrochlore and checkerboard antiferromagnets in a
  strong magnetic field},}\ }\href {\doibase 10.1103/PhysRevB.75.224416}
  {\bibfield  {journal} {\bibinfo  {journal} {Phys. Rev. B}\ }\textbf {\bibinfo
  {volume} {75}},\ \bibinfo {pages} {224416} (\bibinfo {year}
  {2007})}\BibitemShut {NoStop}%
\bibitem [{\citenamefont {Zhitomirsky}\ \emph {et~al.}(2000)\citenamefont
  {Zhitomirsky}, \citenamefont {Honecker},\ and\ \citenamefont
  {Petrenko}}]{zhitomirsky_pyrochlore_2000}%
  \BibitemOpen
  \bibfield  {author} {\bibinfo {author} {\bibfnamefont {M.~E.}\ \bibnamefont
  {Zhitomirsky}}, \bibinfo {author} {\bibfnamefont {A.}~\bibnamefont
  {Honecker}}, \ and\ \bibinfo {author} {\bibfnamefont {O.~A.}\ \bibnamefont
  {Petrenko}},\ }\bibfield  {title} {\enquote {\bibinfo {title} {Field induced
  ordering in highly frustrated antiferromagnets},}\ }\href {\doibase
  10.1103/PhysRevLett.85.3269} {\bibfield  {journal} {\bibinfo  {journal}
  {Phys. Rev. Lett.}\ }\textbf {\bibinfo {volume} {85}},\ \bibinfo {pages}
  {3269--3272} (\bibinfo {year} {2000})}\BibitemShut {NoStop}%
\bibitem [{\citenamefont {Coletta}\ \emph {et~al.}(2013)\citenamefont
  {Coletta}, \citenamefont {Zhitomirsky},\ and\ \citenamefont
  {Mila}}]{coletta_frustrated_2013}%
  \BibitemOpen
  \bibfield  {author} {\bibinfo {author} {\bibfnamefont {Tommaso}\ \bibnamefont
  {Coletta}}, \bibinfo {author} {\bibfnamefont {M.~E.}\ \bibnamefont
  {Zhitomirsky}}, \ and\ \bibinfo {author} {\bibfnamefont {Fr\'ed\'eric}\
  \bibnamefont {Mila}},\ }\bibfield  {title} {\enquote {\bibinfo {title}
  {Quantum stabilization of classically unstable plateau structures},}\ }\href
  {\doibase 10.1103/PhysRevB.87.060407} {\bibfield  {journal} {\bibinfo
  {journal} {Phys. Rev. B}\ }\textbf {\bibinfo {volume} {87}},\ \bibinfo
  {pages} {060407} (\bibinfo {year} {2013})}\BibitemShut {NoStop}%
\bibitem [{\citenamefont {Chen}\ \emph {et~al.}(2018)\citenamefont {Chen},
  \citenamefont {Ran}, \citenamefont {Liu}, \citenamefont {Peng}, \citenamefont
  {Huang},\ and\ \citenamefont {Su}}]{chen_kagome_2018}%
  \BibitemOpen
  \bibfield  {author} {\bibinfo {author} {\bibfnamefont {Xi}~\bibnamefont
  {Chen}}, \bibinfo {author} {\bibfnamefont {Shi-Ju}\ \bibnamefont {Ran}},
  \bibinfo {author} {\bibfnamefont {Tao}\ \bibnamefont {Liu}}, \bibinfo
  {author} {\bibfnamefont {Cheng}\ \bibnamefont {Peng}}, \bibinfo {author}
  {\bibfnamefont {Yi-Zhen}\ \bibnamefont {Huang}}, \ and\ \bibinfo {author}
  {\bibfnamefont {Gang}\ \bibnamefont {Su}},\ }\bibfield  {title} {\enquote
  {\bibinfo {title} {Thermodynamics of spin-1/2 kagomé heisenberg
  antiferromagnet: algebraic paramagnetic liquid and finite-temperature phase
  diagram},}\ }\href {\doibase https://doi.org/10.1016/j.scib.2018.11.007}
  {\bibfield  {journal} {\bibinfo  {journal} {Science Bulletin}\ }\textbf
  {\bibinfo {volume} {63}},\ \bibinfo {pages} {1545--1550} (\bibinfo {year}
  {2018})}\BibitemShut {NoStop}%
\bibitem [{\citenamefont {Nakano}\ and\ \citenamefont
  {Sakai}(2018)}]{nakano_kagome_2018}%
  \BibitemOpen
  \bibfield  {author} {\bibinfo {author} {\bibfnamefont {Hiroki}\ \bibnamefont
  {Nakano}}\ and\ \bibinfo {author} {\bibfnamefont {Tôru}\ \bibnamefont
  {Sakai}},\ }\bibfield  {title} {\enquote {\bibinfo {title}
  {Numerical-diagonalization study of magnetization process of frustrated
  spin-1/2 heisenberg antiferromagnets in two dimensions: —triangular- and
  kagome-lattice antiferromagnets—},}\ }\href {\doibase
  10.7566/JPSJ.87.063706} {\bibfield  {journal} {\bibinfo  {journal} {Journal
  of the Physical Society of Japan}\ }\textbf {\bibinfo {volume} {87}},\
  \bibinfo {pages} {063706} (\bibinfo {year} {2018})}\BibitemShut {NoStop}%
\bibitem [{\citenamefont {Nishimoto}\ \emph
  {et~al.}(2013{\natexlab{b}})\citenamefont {Nishimoto}, \citenamefont
  {Shibata},\ and\ \citenamefont {Hotta}}]{nishimoto_controlling_2013}%
  \BibitemOpen
  \bibfield  {author} {\bibinfo {author} {\bibfnamefont {Satoshi}\ \bibnamefont
  {Nishimoto}}, \bibinfo {author} {\bibfnamefont {Naokazu}\ \bibnamefont
  {Shibata}}, \ and\ \bibinfo {author} {\bibfnamefont {Chisa}\ \bibnamefont
  {Hotta}},\ }\bibfield  {title} {\enquote {\bibinfo {title} {Controlling
  frustrated liquids and solids with an applied field in a kagome {Heisenberg}
  antiferromagnet},}\ }\href {\doibase 10.1038/ncomms3287} {\bibfield
  {journal} {\bibinfo  {journal} {Nature Communications}\ }\textbf {\bibinfo
  {volume} {4}},\ \bibinfo {pages} {2287} (\bibinfo {year}
  {2013}{\natexlab{b}})}\BibitemShut {NoStop}%
\bibitem [{\citenamefont {Plumb}\ \emph {et~al.}(2019)\citenamefont {Plumb},
  \citenamefont {Changlani}, \citenamefont {Scheie}, \citenamefont {Zhang},
  \citenamefont {Krizan}, \citenamefont {Rodriguez-Rivera}, \citenamefont
  {Qiu}, \citenamefont {Winn}, \citenamefont {Cava},\ and\ \citenamefont
  {Broholm}}]{plumb_continuum_2019}%
  \BibitemOpen
  \bibfield  {author} {\bibinfo {author} {\bibfnamefont {K.~W.}\ \bibnamefont
  {Plumb}}, \bibinfo {author} {\bibfnamefont {Hitesh~J.}\ \bibnamefont
  {Changlani}}, \bibinfo {author} {\bibfnamefont {A.}~\bibnamefont {Scheie}},
  \bibinfo {author} {\bibfnamefont {Shu}\ \bibnamefont {Zhang}}, \bibinfo
  {author} {\bibfnamefont {J.~W.}\ \bibnamefont {Krizan}}, \bibinfo {author}
  {\bibfnamefont {J.~A.}\ \bibnamefont {Rodriguez-Rivera}}, \bibinfo {author}
  {\bibfnamefont {Yiming}\ \bibnamefont {Qiu}}, \bibinfo {author}
  {\bibfnamefont {B.}~\bibnamefont {Winn}}, \bibinfo {author} {\bibfnamefont
  {R.~J.}\ \bibnamefont {Cava}}, \ and\ \bibinfo {author} {\bibfnamefont
  {C.~L.}\ \bibnamefont {Broholm}},\ }\bibfield  {title} {\enquote {\bibinfo
  {title} {Continuum of quantum fluctuations in a three-dimensional {S} = 1
  {Heisenberg} magnet},}\ }\href {\doibase 10.1038/s41567-018-0317-3}
  {\bibfield  {journal} {\bibinfo  {journal} {Nature Physics}\ }\textbf
  {\bibinfo {volume} {15}},\ \bibinfo {pages} {54--59} (\bibinfo {year}
  {2019})}\BibitemShut {NoStop}%
\bibitem [{\citenamefont {Hubig}\ \emph {et~al.}()\citenamefont {Hubig},
  \citenamefont {Lachenmaier}, \citenamefont {Linden}, \citenamefont
  {Reinhard}, \citenamefont {Stenzel}, \citenamefont {Swoboda},\ and\
  \citenamefont {Grundner}}]{hubig:_syten_toolk}%
  \BibitemOpen
  \bibfield  {author} {\bibinfo {author} {\bibfnamefont {Claudius}\
  \bibnamefont {Hubig}}, \bibinfo {author} {\bibfnamefont {Felix}\ \bibnamefont
  {Lachenmaier}}, \bibinfo {author} {\bibfnamefont {Nils-Oliver}\ \bibnamefont
  {Linden}}, \bibinfo {author} {\bibfnamefont {Teresa}\ \bibnamefont
  {Reinhard}}, \bibinfo {author} {\bibfnamefont {Leo}\ \bibnamefont {Stenzel}},
  \bibinfo {author} {\bibfnamefont {Andreas}\ \bibnamefont {Swoboda}}, \ and\
  \bibinfo {author} {\bibfnamefont {Martin}\ \bibnamefont {Grundner}},\ }\href
  {https://syten.eu} {\enquote {\bibinfo {title} {The \textsc{SyTen}
  toolkit},}\ }\BibitemShut {NoStop}%
\bibitem [{\citenamefont {Hubig}(2017)}]{hubig17:_symmet_protec_tensor_networ}%
  \BibitemOpen
  \bibfield  {author} {\bibinfo {author} {\bibfnamefont {Claudius}\
  \bibnamefont {Hubig}},\ }\emph {\bibinfo {title} {Symmetry-Protected Tensor
  Networks}},\ \href {https://edoc.ub.uni-muenchen.de/21348/} {Ph.D. thesis},\
  \bibinfo  {school} {LMU M\"unchen} (\bibinfo {year} {2017})\BibitemShut
  {NoStop}%
\bibitem [{\citenamefont {Hubig}\ \emph {et~al.}(2015)\citenamefont {Hubig},
  \citenamefont {McCulloch}, \citenamefont {Schollw\"ock},\ and\ \citenamefont
  {Wolf}}]{hubig_2015}%
  \BibitemOpen
  \bibfield  {author} {\bibinfo {author} {\bibfnamefont {C.}~\bibnamefont
  {Hubig}}, \bibinfo {author} {\bibfnamefont {I.~P.}\ \bibnamefont
  {McCulloch}}, \bibinfo {author} {\bibfnamefont {U.}~\bibnamefont
  {Schollw\"ock}}, \ and\ \bibinfo {author} {\bibfnamefont {F.~A.}\
  \bibnamefont {Wolf}},\ }\bibfield  {title} {\enquote {\bibinfo {title}
  {Strictly single-site dmrg algorithm with subspace expansion},}\ }\href
  {\doibase 10.1103/PhysRevB.91.155115} {\bibfield  {journal} {\bibinfo
  {journal} {Phys. Rev. B}\ }\textbf {\bibinfo {volume} {91}},\ \bibinfo
  {pages} {155115} (\bibinfo {year} {2015})}\BibitemShut {NoStop}%
\bibitem [{\citenamefont {McCulloch}(2007)}]{McCulloch_2007}%
  \BibitemOpen
  \bibfield  {author} {\bibinfo {author} {\bibfnamefont {Ian~P}\ \bibnamefont
  {McCulloch}},\ }\bibfield  {title} {\enquote {\bibinfo {title} {From
  density-matrix renormalization group to matrix product states},}\ }\href
  {\doibase 10.1088/1742-5468/2007/10/p10014} {\bibfield  {journal} {\bibinfo
  {journal} {Journal of Statistical Mechanics: Theory and Experiment}\ }\textbf
  {\bibinfo {volume} {2007}},\ \bibinfo {pages} {P10014--P10014} (\bibinfo
  {year} {2007})}\BibitemShut {NoStop}%
\bibitem [{\citenamefont {White}(1992)}]{white_1992}%
  \BibitemOpen
  \bibfield  {author} {\bibinfo {author} {\bibfnamefont {Steven~R.}\
  \bibnamefont {White}},\ }\bibfield  {title} {\enquote {\bibinfo {title}
  {Density matrix formulation for quantum renormalization groups},}\ }\href
  {\doibase 10.1103/PhysRevLett.69.2863} {\bibfield  {journal} {\bibinfo
  {journal} {Phys. Rev. Lett.}\ }\textbf {\bibinfo {volume} {69}},\ \bibinfo
  {pages} {2863--2866} (\bibinfo {year} {1992})}\BibitemShut {NoStop}%
\bibitem [{\citenamefont {White}(1993)}]{white_1993}%
  \BibitemOpen
  \bibfield  {author} {\bibinfo {author} {\bibfnamefont {Steven~R.}\
  \bibnamefont {White}},\ }\bibfield  {title} {\enquote {\bibinfo {title}
  {Density-matrix algorithms for quantum renormalization groups},}\ }\href
  {\doibase 10.1103/PhysRevB.48.10345} {\bibfield  {journal} {\bibinfo
  {journal} {Phys. Rev. B}\ }\textbf {\bibinfo {volume} {48}},\ \bibinfo
  {pages} {10345--10356} (\bibinfo {year} {1993})}\BibitemShut {NoStop}%
\bibitem [{\citenamefont {Noack}\ \emph {et~al.}(2005)\citenamefont {Noack},
  \citenamefont {Manmana}, \citenamefont {Avella},\ and\ \citenamefont
  {Mancini}}]{noack2005}%
  \BibitemOpen
  \bibfield  {author} {\bibinfo {author} {\bibfnamefont {Reinhard~M.}\
  \bibnamefont {Noack}}, \bibinfo {author} {\bibfnamefont {Salvatore~R.}\
  \bibnamefont {Manmana}}, \bibinfo {author} {\bibfnamefont {Adolfo}\
  \bibnamefont {Avella}}, \ and\ \bibinfo {author} {\bibfnamefont {Ferdinando}\
  \bibnamefont {Mancini}},\ }\bibfield  {title} {\enquote {\bibinfo {title}
  {Diagonalization‐ and numerical renormalization‐group‐based methods for
  interacting quantum systems},}\ }\href {\doibase 10.1063/1.2080349}
  {\bibfield  {journal} {\bibinfo  {journal} {AIP Conference Proceedings}\
  }\textbf {\bibinfo {volume} {789}},\ \bibinfo {pages} {93--163} (\bibinfo
  {year} {2005})}\BibitemShut {NoStop}%
\bibitem [{\citenamefont {Schollw\"ock}(2011)}]{schollwock_review_2011}%
  \BibitemOpen
  \bibfield  {author} {\bibinfo {author} {\bibfnamefont {Ulrich}\ \bibnamefont
  {Schollw\"ock}},\ }\bibfield  {title} {\enquote {\bibinfo {title} {The
  density-matrix renormalization group in the age of matrix product states},}\
  }\href {\doibase https://doi.org/10.1016/j.aop.2010.09.012} {\bibfield
  {journal} {\bibinfo  {journal} {Annals of Physics}\ }\textbf {\bibinfo
  {volume} {326}},\ \bibinfo {pages} {96 -- 192} (\bibinfo {year}
  {2011})}\BibitemShut {NoStop}%
\bibitem [{\citenamefont {Hallberg}(2006)}]{hallberg_review}%
  \BibitemOpen
  \bibfield  {author} {\bibinfo {author} {\bibfnamefont {Karen~A.}\
  \bibnamefont {Hallberg}},\ }\bibfield  {title} {\enquote {\bibinfo {title}
  {New trends in density matrix renormalization},}\ }\href {\doibase
  10.1080/00018730600766432} {\bibfield  {journal} {\bibinfo  {journal}
  {Advances in Physics}\ }\textbf {\bibinfo {volume} {55}},\ \bibinfo {pages}
  {477--526} (\bibinfo {year} {2006})}\BibitemShut {NoStop}%
\bibitem [{\citenamefont {Stoudenmire}\ and\ \citenamefont
  {White}(2012)}]{white_2d_dmrg}%
  \BibitemOpen
  \bibfield  {author} {\bibinfo {author} {\bibfnamefont {E.M.}\ \bibnamefont
  {Stoudenmire}}\ and\ \bibinfo {author} {\bibfnamefont {Steven~R.}\
  \bibnamefont {White}},\ }\bibfield  {title} {\enquote {\bibinfo {title}
  {Studying two-dimensional systems with the density matrix renormalization
  group},}\ }\href {\doibase 10.1146/annurev-conmatphys-020911-125018}
  {\bibfield  {journal} {\bibinfo  {journal} {Annual Review of Condensed Matter
  Physics}\ }\textbf {\bibinfo {volume} {3}},\ \bibinfo {pages} {111--128}
  (\bibinfo {year} {2012})}\BibitemShut {NoStop}%
\bibitem [{\citenamefont {Ummethum}\ \emph {et~al.}(2013)\citenamefont
  {Ummethum}, \citenamefont {Schnack},\ and\ \citenamefont
  {L{\"a}uchli}}]{ummethum_numerics_2013}%
  \BibitemOpen
  \bibfield  {author} {\bibinfo {author} {\bibfnamefont {J.}~\bibnamefont
  {Ummethum}}, \bibinfo {author} {\bibfnamefont {J.}~\bibnamefont {Schnack}}, \
  and\ \bibinfo {author} {\bibfnamefont {A.~M.}\ \bibnamefont {L{\"a}uchli}},\
  }\bibfield  {title} {\enquote {\bibinfo {title} {Large-scale numerical
  investigations of the antiferromagnetic heisenberg icosidodecahedron},}\
  }\href {\doibase https://doi.org/10.1016/j.jmmm.2012.09.037} {\bibfield
  {journal} {\bibinfo  {journal} {Journal of Magnetism and Magnetic Materials}\
  }\textbf {\bibinfo {volume} {327}},\ \bibinfo {pages} {103 -- 109} (\bibinfo
  {year} {2013})}\BibitemShut {NoStop}%
\bibitem [{\citenamefont {Hubig}\ \emph {et~al.}(2018)\citenamefont {Hubig},
  \citenamefont {Haegeman},\ and\ \citenamefont
  {Schollw\"ock}}]{hubig_prb_2018}%
  \BibitemOpen
  \bibfield  {author} {\bibinfo {author} {\bibfnamefont {C.}~\bibnamefont
  {Hubig}}, \bibinfo {author} {\bibfnamefont {J.}~\bibnamefont {Haegeman}}, \
  and\ \bibinfo {author} {\bibfnamefont {U.}~\bibnamefont {Schollw\"ock}},\
  }\bibfield  {title} {\enquote {\bibinfo {title} {Error estimates for
  extrapolations with matrix-product states},}\ }\href {\doibase
  10.1103/PhysRevB.97.045125} {\bibfield  {journal} {\bibinfo  {journal} {Phys.
  Rev. B}\ }\textbf {\bibinfo {volume} {97}},\ \bibinfo {pages} {045125}
  (\bibinfo {year} {2018})}\BibitemShut {NoStop}%
\bibitem [{\citenamefont {Vanderstraeten}\ \emph {et~al.}(2018)\citenamefont
  {Vanderstraeten}, \citenamefont {Vanhecke},\ and\ \citenamefont
  {Verstraete}}]{vanderstraeten_residual_2018}%
  \BibitemOpen
  \bibfield  {author} {\bibinfo {author} {\bibfnamefont {Laurens}\ \bibnamefont
  {Vanderstraeten}}, \bibinfo {author} {\bibfnamefont {Bram}\ \bibnamefont
  {Vanhecke}}, \ and\ \bibinfo {author} {\bibfnamefont {Frank}\ \bibnamefont
  {Verstraete}},\ }\bibfield  {title} {\enquote {\bibinfo {title} {Residual
  entropies for three-dimensional frustrated spin systems with tensor
  networks},}\ }\href {\doibase 10.1103/PhysRevE.98.042145} {\bibfield
  {journal} {\bibinfo  {journal} {Physical Review E}\ }\textbf {\bibinfo
  {volume} {98}},\ \bibinfo {pages} {042145} (\bibinfo {year}
  {2018})}\BibitemShut {NoStop}%
\bibitem [{\citenamefont {Nagle}(1966{\natexlab{a}})}]{nagle_lattice_1966}%
  \BibitemOpen
  \bibfield  {author} {\bibinfo {author} {\bibfnamefont {J.~F.}\ \bibnamefont
  {Nagle}},\ }\bibfield  {title} {\enquote {\bibinfo {title} {Lattice
  {Statistics} of {Hydrogen} {Bonded} {Crystals}. {I}. {The} {Residual}
  {Entropy} of {Ice}},}\ }\href {\doibase 10.1063/1.1705058} {\bibfield
  {journal} {\bibinfo  {journal} {Journal of Mathematical Physics}\ }\textbf
  {\bibinfo {volume} {7}},\ \bibinfo {pages} {1484--1491} (\bibinfo {year}
  {1966}{\natexlab{a}})}\BibitemShut {NoStop}%
\bibitem [{\citenamefont
  {Nagle}(1966{\natexlab{b}})}]{nagle_lattice_1966_diamond_lattice}%
  \BibitemOpen
  \bibfield  {author} {\bibinfo {author} {\bibfnamefont {John~F.}\ \bibnamefont
  {Nagle}},\ }\bibfield  {title} {\enquote {\bibinfo {title} {New
  series-expansion method for the dimer problem},}\ }\href {\doibase
  10.1103/PhysRev.152.190} {\bibfield  {journal} {\bibinfo  {journal} {Phys.
  Rev.}\ }\textbf {\bibinfo {volume} {152}},\ \bibinfo {pages} {190--197}
  (\bibinfo {year} {1966}{\natexlab{b}})}\BibitemShut {NoStop}%
\bibitem [{\citenamefont {{Moessner}}(1998)}]{Moessner_relief}%
  \BibitemOpen
  \bibfield  {author} {\bibinfo {author} {\bibfnamefont {R.}~\bibnamefont
  {{Moessner}}},\ }\bibfield  {title} {\enquote {\bibinfo {title} {{Relief and
  generation of frustration in pyrochlore magnets by single-ion anisotropy}},}\
  }\href {\doibase 10.1103/PhysRevB.57.R5587} {\bibfield  {journal} {\bibinfo
  {journal} {\prb}\ }\textbf {\bibinfo {volume} {57}},\ \bibinfo {pages}
  {R5587--R5589} (\bibinfo {year} {1998})}\BibitemShut {NoStop}%
\bibitem [{\citenamefont {Hubig}\ \emph {et~al.}(2017)\citenamefont {Hubig},
  \citenamefont {McCulloch},\ and\ \citenamefont
  {Schollw\"ock}}]{hubig_mpo_2017}%
  \BibitemOpen
  \bibfield  {author} {\bibinfo {author} {\bibfnamefont {C.}~\bibnamefont
  {Hubig}}, \bibinfo {author} {\bibfnamefont {I.~P.}\ \bibnamefont
  {McCulloch}}, \ and\ \bibinfo {author} {\bibfnamefont {U.}~\bibnamefont
  {Schollw\"ock}},\ }\bibfield  {title} {\enquote {\bibinfo {title} {Generic
  construction of efficient matrix product operators},}\ }\href {\doibase
  10.1103/PhysRevB.95.035129} {\bibfield  {journal} {\bibinfo  {journal} {Phys.
  Rev. B}\ }\textbf {\bibinfo {volume} {95}},\ \bibinfo {pages} {035129}
  (\bibinfo {year} {2017})}\BibitemShut {NoStop}%
\bibitem [{\citenamefont {Zheng}\ \emph {et~al.}(2017)\citenamefont {Zheng},
  \citenamefont {Chung}, \citenamefont {Corboz}, \citenamefont {Ehlers},
  \citenamefont {Qin}, \citenamefont {Noack}, \citenamefont {Shi},
  \citenamefont {White}, \citenamefont {Zhang},\ and\ \citenamefont
  {Chan}}]{doi:10.1126/science.aam7127}%
  \BibitemOpen
  \bibfield  {author} {\bibinfo {author} {\bibfnamefont {Bo-Xiao}\ \bibnamefont
  {Zheng}}, \bibinfo {author} {\bibfnamefont {Chia-Min}\ \bibnamefont {Chung}},
  \bibinfo {author} {\bibfnamefont {Philippe}\ \bibnamefont {Corboz}}, \bibinfo
  {author} {\bibfnamefont {Georg}\ \bibnamefont {Ehlers}}, \bibinfo {author}
  {\bibfnamefont {Ming-Pu}\ \bibnamefont {Qin}}, \bibinfo {author}
  {\bibfnamefont {Reinhard~M.}\ \bibnamefont {Noack}}, \bibinfo {author}
  {\bibfnamefont {Hao}\ \bibnamefont {Shi}}, \bibinfo {author} {\bibfnamefont
  {Steven~R.}\ \bibnamefont {White}}, \bibinfo {author} {\bibfnamefont
  {Shiwei}\ \bibnamefont {Zhang}}, \ and\ \bibinfo {author} {\bibfnamefont
  {Garnet Kin-Lic}\ \bibnamefont {Chan}},\ }\bibfield  {title} {\enquote
  {\bibinfo {title} {Stripe order in the underdoped region of the
  two-dimensional hubbard model},}\ }\href {\doibase 10.1126/science.aam7127}
  {\bibfield  {journal} {\bibinfo  {journal} {Science}\ }\textbf {\bibinfo
  {volume} {358}},\ \bibinfo {pages} {1155--1160} (\bibinfo {year}
  {2017})}\BibitemShut {NoStop}%
\end{thebibliography}%
\clearpage
\appendix

\section{Further details of the DMRG simulation}\label{sec:numerical}

Matrix-product operators (MPOs) can be constructed by hand for the simplest 1D models with nearest-neighbor interactions. However, this task becomes difficult when long-range interactions are present and needs to be done in an automatic way. While the corresponding MPO of a single product, e.g. $S_i^zS_j^z$, can be represented by an MPO of small bond dimension, the bond dimension grows exponentially by summing up multiple terms. Luckily, the MPO can be compressed effectively using the deparallelisation method \cite{hubig_mpo_2017} without any information loss in many cases. We start by optimizing a random matrix-product state (MPS) with the corresponding size of the cluster using DMRG where the long-range correlations are captured automatically up to a given bond dimension. The choice of the 3D$\rightarrow$1D mapping can have an influence to the overall convergence as well as the final bond dimension of the MPO, which can also have an impact on the speedup of the calculation. However, we did not observe a significant difference in terms of computation time and convergence properties for the different paths in the three dimensional clusters we investigated.

The convergence problems in three dimensions are even more severe than in two dimensions, since there are more periodic bonds yielding to increasing long range interaction. While we do not face convergence problems for the 32 and 48 site clusters, DMRG often exhibits bad convergence and gets stuck
in local minima for the larger clusters due to its local-update nature.
In these cases, the pattern in Fig.~\ref{fig:magnetization} is only partially generated if we initialize the algorithm with random states. This can be monitored either by examining the truncation error or the two-site variance as a function of the bond dimension. Metastable states induce a non-monotonic behavior of these quantities, that is, the energy decreases but the truncation error or two-site variance increases. When the convergence is smooth, the energy follows typically a linear behavior as a function of the two-site variance \cite{hubig_prb_2018}.

\begin{figure}
    \centering
\includegraphics[width=\columnwidth]{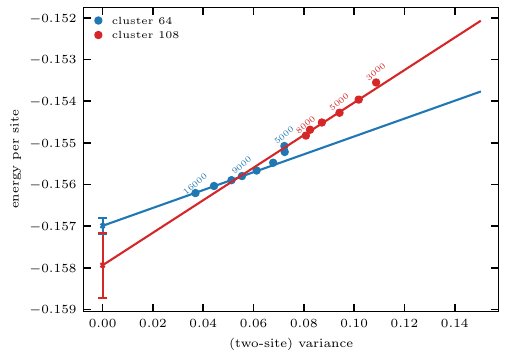}
    \caption{The energy extrapolations (solid lines) for the ${1}/{2}$ plateau state of the 64 and 108 site clusters as a function of the two-site variance. The labels indicate the corresponding U(1) bond dimensions. The last four points with the smallest variance are included in the linear fits.}
        \label{fig:extrapolation}
\end{figure}

To stabilize one of the magnetically ordered states in our simulation, we therefore use the pinning-field technique \cite{white_2d_dmrg}. 
We apply a magnetic field (at low bond dimensions) to the same set of sites for each tetrahedral unit cell which compatible with the polarized kagom\'{e} planes such that the ordered state is stabilized. We then remove this pinning field and continue the optimization procedure while further increasing the bond dimension. 
With this approach the overall convergence becomes much better and smoother as it is indicated by Fig.~\ref{fig:extrapolation}. 

{
    To assess the quality of the variational ground state, we compare the total energies at a fixed bond dimension $\chi=8000$ for the 64 and 108 site clusters using different initial conditions: starting from random initial states or using the pinning-field technique to start from ordered patterns.
	The ordered pattern produces lower energies ($\sim 0.1 J$) than starting from random initial states.
We also consider another possible pattern, the $R$-state \cite{balents_pyrochlore_2006}, which was predicted for the $S=3/2$ case. 
This state also yields higher energies than our best variational state, although only with a small difference $\sim 0.05J$ and is therefore clearly in the low energy manifold. }

The error of the extrapolated energies are defined as one quarter of the distance between the lowest energy DMRG datapoint and the extrapolated value, which is commonly used in the community \cite{doi:10.1126/science.aam7127}. The same definition of the error bars is used in the main text as well and should be understood as an estimate of the systemic extrapolation error.

\section{Localized magnons in high fields}\label{sec:magnon}

Constructing analytic solutions for quantum many body systems is a 
challenging discipline and succeeds only in special cases. Therefore, solutions of 
the ground state in the form of quasi particles in high fields were a huge success for spin systems in certain frustrated lattices \cite{schollwoeck_qm_2004, honecker_magnon_2004, 
schnack_eigenstates_2006, derzhko_localized_magnons_2007, 
schulenburg_theory_kagome_magnon_groundstate_2002, 
nishimoto_numerics_kagome_plateaus_2013,honecker_kagome_2005}. Probably the most famous example for 
these quasi particles are independent and localized magnon excitations in the 
two dimensional kagom\'{e} lattice.
\paragraph*{Kagom\'{e} lattice}
Localized magnon excitations describe the ground state of the Heisenberg model in a large external field on the kagom\'{e} lattice. Theses are confined to non-touching hexagons such that the description can be limited to a single star of David. Starting from the fully polarized state $\vert \uparrow 
...\uparrow\rangle$, a single magnon state is given by (up to normalization)
\begin{align}
	\vert m\rangle \propto \sum_{j\in\varhexagon} (-1)^j S^-_j\vert \uparrow 
...\uparrow\rangle\label{eq:mag_00}
\end{align}
where $j$ runs over the hexagon. The localization can be easily verified since each 
corner spin of the star is attached to two spins of the inner hexagon. The contributions of 
flipped spins propagating to corner sites are canceled due the alternating 
sign. {Hence, the magnetic excitation remains within the hexagon and preserves the alternating sign structure such that the hopping contribution is diagonal}. Not only are the magnons localized but also they are ground states of the 
Heisenberg antiferromagnet for high fields. For simplification we set $h=0$ and 
focus on the hopping $H_\pm$ and interaction term $H_z$ individually:
\begin{align}
	H &= JH_{\pm} + JH_z\nonumber\\
	&= \frac{J}{2}\sum_{\langle i, j\rangle} \left[ \sigma^+_i\sigma^-_j + 
\sigma^-_i\sigma^+_j\right] + \frac{J}{4} \sum_{\langle i, j\rangle} \sigma^z_i 
\sigma^z_j\label{eq:mag_01}
\end{align}
Within the hexagon, the sign of each term in Eq. \eqref{eq:mag_00} is inverted by 
$H_\pm$. Hence, $H_\pm\vert m\rangle = -J\vert m\rangle${ and the hopping term reduces the energy by $J$}. In the kagom\'{e} lattice, each site is 
attached to 4 other sites. The contribution to the energy by the interaction 
$H_z$ due to single spin flip is a reduction of $2J$ in contrast to the fully 
polarized state. In total, a single magnon reduces the energy by $3J$.\\
Due to the localization, multiple independent states can be placed within the 
kagom\'{e} lattice as long as they are separated by at least one site. In this 
manner, each hexagon is uniquely assigned to 9 sites in the kagom\'{e} lattice to 
achieve the densest filling. The complete tiling of hexagons describes the 
ground state and corresponds to a magnetization of $m/m_{\text{sat}} = 7/9$. The 
energy per site is reduced to $E_{7/9} = 1/6J$ from the fully polarized state 
$E_1= 1/2J$.
\begin{figure}
    \centering
\includegraphics[width=\columnwidth]{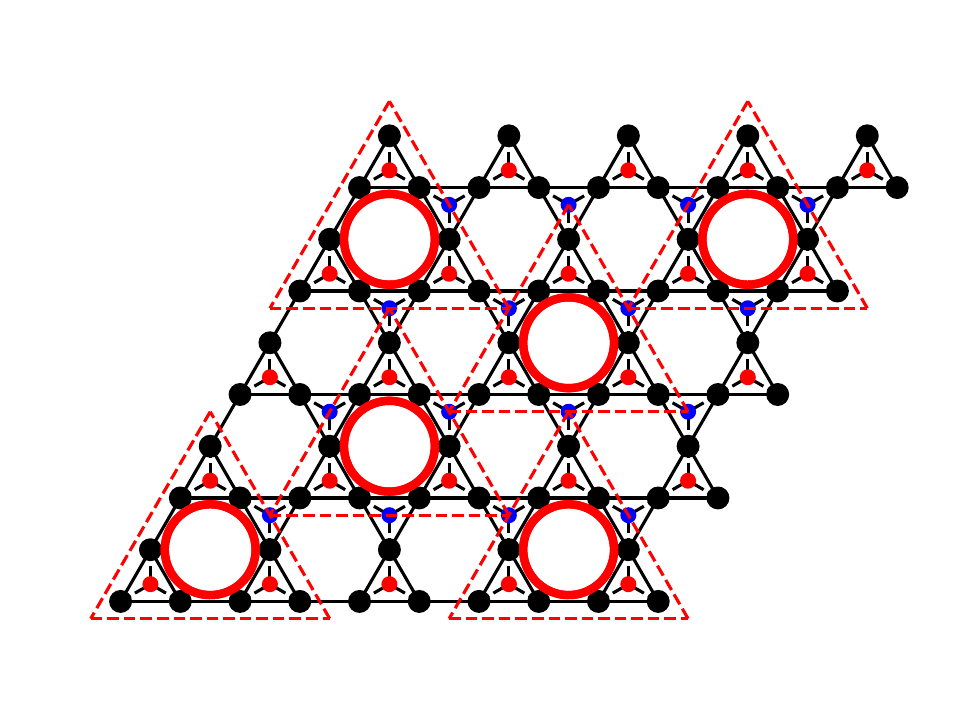}
    \caption{Complete Hexagon tiling in a single kagom\'{e} plane (black sites) of 
the pyrochlore lattice. Red and blue sites correspond to the upper and lower 
separating layer respectively. Possible localized magnon states on the 
corresponding hexagons are illustrated with red circles. All sites are uniquely 
assigned to one localized hexagon within the red triangles.}
        \label{fig:mag_00}
\end{figure}

\paragraph*{Pyrochlore lattice}
The tetrahedral unit cell of the pyrochlore lattice consist of four sites. Each 
face of this tetrahedron defines one out of four orientations of parallel kagom\'{e} planes in the lattice. The apex spins form a separating triangular plane between the kagom\'{e} planes. Equivalently to the two dimensional case, the same formalism can be used to generate localized magnons in the pyrochlore 
lattice. As visualized in Fig. \ref{fig:mag_00} by the red dotted triangles, 9 
sites are uniquely assigned to each localized hexagon in the kagom\'{e} plane (black 
sites), such that no supercells are sharing the same site. The corresponding 
magnon excitation is illustrated by the red circles. Red and blue sites 
correspond to the upper and lower separating triangular layer respectively. In addition to 
the 9 sites laying inside the plane, we include the 3 (red) sites from the upper 
layer to realize a complete tiling of the pyrochlore lattice. As in the pure 
two dimensional case, a localized magnon is confined to 3 unit cells. Hence, 
each magnon is assigned to 12 sites and the corresponding plateau is 
$m/m_{\text{sat}} = 5/6$.

We can determine the saturation field analytically by comparing the energy per site of the fully polarized state, $E_1=\frac{3}{4}J-\frac{h}{2}$, with the energy of the magnon crystal, $E_{5/6} = \frac{5}{12}J - \frac{5}{6}\frac{h}{2}$.
The first part is derived from the Heisenberg Hamiltonian in Eq. \eqref{eq:mag_00} and the second part is the energy shift induced by the external field.
\begin{align}
	& & E_1(h_\text{sat}) & = E_{5/6}(h_\text{sat}) & &\\\label{eq:mag_02}
	& & \frac{3}{4}J - \frac{h_\text{sat}}{2} & = \frac{5}{12}J - \frac{5}{6}\frac{h_\text{sat}}{2} & &\nonumber\\
	\Rightarrow & & h_\text{sat} = 4J& & &
\end{align}


\end{document}